\documentclass[10pt,final,journal,oneside,twocolumn]{IEEEtran}

\usepackage{psfrag}
\usepackage{subfigure}
 \usepackage{flushend}
\usepackage{epsfig}
\usepackage[usenames]{color}
\usepackage[tbtags,sumlimits,nointlimits,reqno]{amsmath}
\usepackage{amssymb}
\usepackage{cite}
\usepackage[super]{nth}

\newcommand{\red}[1]{\textcolor{black}{#1}}
\newcommand{\ok}[1]{#1}

\newcommand{\tjs}[1]{#1}

 \newcommand{\sg}[1]{\textcolor{black}{#1}}

\usepackage{gensymb}
\usepackage{cite} 
\usepackage{overpic}
\usepackage{cuted}

\usepackage{esint}

\renewcommand{\r}{\mathbf{r}}

\newcommand{\E}{\mathbf{E}}
\newcommand{\J}{\mathbf{J}}
\renewcommand{\H}{\mathbf{H}}

\newcommand{\X}{\mathbf{X}}
\newcommand{\M}{\mathbf{M}}

\newcommand{\TT}{\mathbf{T}}
\newcommand{\GGamma}{\mathbf{\Gamma}}
\newcommand{\barI}{\bar{\mathbf{I}}}
\newcommand{\uhat}{\hat{\mathbf{u}}}
\newcommand{\khat}{\hat{\mathbf{k}}}

\newcommand{\nhat}{\hat{\mathbf{n}}}
\newcommand{\rhat}{\hat{\mathbf{r}}}

\newcommand{\q}{\mathbf{q}}
\newcommand{\p}{\mathbf{p}}

\renewcommand{\P}{\mathbf{P}}
\newcommand{\G}{\mathbf{G}}
\newcommand{\A}{\mathbf{A}}

 \definecolor{amber}{rgb}{0.8, 0.33, 0.0}
 \definecolor{almond}{rgb}{0.94, 0.87, 0.8}

\usepackage{tikz}
\usetikzlibrary{shapes.geometric, arrows,automata,positioning,shapes.multipart}
\usepackage{tabularx,colortbl}
\usetikzlibrary{fit,backgrounds} % <-added
\usepackage{bm}
\usepackage{verbatim}

\begin{document}

% \title{RO-GSTC Ray Tracer for Electrically Large Metasurfaces using Surface Susceptibilities - Part II}
\title{Ray-Optical Evaluation of Scattering from  \sg{Electrically} Large Metasurfaces Characterized by Locally Periodic \ok{Surface Susceptibilities}}

\author{Scott Stewart, Yvo L. C. de Jong,~\IEEEmembership{Senior Member,~IEEE}, Tom J. Smy\\ and Shulabh Gupta,~\IEEEmembership{Senior Member,~IEEE}

\thanks{Scott Stewart, T. J. Smy and S.~Gupta, are with the Department of Electronics, Carleton University, Ottawa, Ontario, Canada. Email: scott.stewart@carleton.ca.}
\thanks{Y. L. C. de Jong is with the Communications Research Centre Canada (CRC), Ottawa, Ontario, Canada. Email: yvo.dejong@ised-isde.gc.ca.}
}
 %The paper headers
\markboth{MANUSCRIPT DRAFT}
{Shell \MakeLowercase{\textit{et al.}}: Bare Demo of IEEEtran.cls for Journals}

% make the title area
\maketitle

\begin{abstract}
This work continues the development of the ray-tracing method of \cite{yvo}
%based on geometrical optics (GO) and the geometrical theory of diffraction (GTD) \tjs{specified by} transmission and reflection coefficients $\{ \mathbf{\Gamma}, \T\}$
for computing the scattered fields from metasurfaces \ok{characterized by locally periodic reflection and transmission coefficients.}
In this work, instead of describing the metasurface in terms of \ok{scattering coefficients that depend on the incidence direction,}
%\tjs{quantities} $\{\mathbf{\Gamma}, \T\}$,
%the metasurface
\ok{its} scattering behavior is \ok{characterized by the surface susceptibility tensors that appear in the generalized sheet transition conditions (GSTCs).}
\ok{As the latter quantities are} constitutive parameters, \ok{they do not depend on the incident field and thus enable a more} compact and physically motivated description of the surface.
%i.e. angle independent surface susceptibilities $\bar{\bm\chi}$, leading to \tjs{a} compact and physically motivated description of the surface.
%For slowly varying non-uniform metasurfaces, the
The locally periodic susceptibility profile is expanded into a Fourier series, and the GSTCs \ok{are rewritten in a form that enables them to be numerically solved for} \red{in terms of} the \ok{reflected and transmitted surface} fields.
\ok{The scattered field at arbitrary detector locations is constructed by evaluating critical-point contributions of the first and second kinds using a Forward Ray Tracing (FRT) scheme.}
%Using a Forward Ray Tracing (FRT) scheme based on diffraction coefficients in \cite{yvo}, the total fields are constructed anywhere in the region using critical points contributions of various kinds.
The \ok{accuracy of the} resulting framework
%is next compared and
\ok{has been} verified with \ok{an} Integral Equation based \sg{Boundary Element Method (BEM)-}GSTC full-wave solver for a variety of examples such as \ok{a} periodically modulated metasurface, \ok{a} metasurface diffuser and \ok{a} beam collimator.
\end{abstract}


%\tableofcontents

\section{Introduction}

\IEEEPARstart{E}{lectromagnetic} (EM) metasurfaces have recently emerged as spatial field processors,
%where they
\ok{which} can be engineered to transform incident wavefronts into desired wavefronts in both reflection and transmission. They are thin surfaces
%($\delta\ll \lambda_0$)
\ok{(with thicknesses much smaller than the wavelength, $\lambda$)} consisting of an array of sub-wavelength scatterers \ok{(or particles)} which interact with the incident waves, and through their rich electric and magnetic polarization responses, can manipulate their amplitude, phase and polarizations. Consequently, they have found \tjs{a} myriad of applications across the entire electromagnetic spectrum ranging from optics to microwaves \cite{MetaHolo,MetaCloak,MetaFieldTransformation,ReconfgMSoptics,ShaltoutSTMetasurface}.

While they are created using engineered sub-wavelength unit-cell structures, practical metasurfaces are electrically large \tjs{--} of the order of several tens of wavelengths or more.
The incident
%waves
\ok{field} interacts with the non-uniformly arrayed resonating particles to generate
%the microscopic response
\ok{a spatial distribution of localized} microscopic responses \ok{which together}  \sg{with} the edge \ok{effects due to the finite size of the surface} aperture, produce the overall macroscopic response. In addition, if the metasurface is part of \ok{a} larger system \ok{of scattering objects}, the
\ok{field interactions}
%scattered fields from the metasurface
become even more complex and the \ok{scattering} problem
%of field scattering from metasurfaces
\tjs{must} be seen as a multi-scale field problem -- i.e., from the sub-wavelength unit-cell level to the system level.

The \ok{implications of the} multi-scale nature of the metasurface scattering problem may be \ok{better} understood by considering the application \ok{of metasurfaces} to optimizing wireless communication performance at microwave and mmWave frequencies.
For instance, radio-frequency (RF) metasurfaces are envisioned to act as smart scatterers manipulating EM wave propagation in a given radio environment, avoiding obstacles or redirecting
%waves
\ok{wireless signals} to desired receiver locations using the best possible channel path \cite{Fink_AI_Metasurface, Basr_MIMO6G, Zhang_Smart_MS_WirelessComm}. Another application to consider is that of EM illusions and camouflaging \cite{smy2020surface, smy2020surface_Camouflage}\tjs{, where }electrically large metasurfaces can act as holograms to project a virtual image of an object, or to enclose a real object, camouflaging it against a complex scattering environment. In all cases, such environments are naturally electrically large, with a variety of static and moving objects, of sizes which are orders of magnitude larger than those of the metasurface unit cells. 

The problem of determining the scattered fields from an electrically large surface, possibly in the presence of other scattering objects, is
%thus
a challenging task.
\ok{Among the variety of numerical methods} that have recently been proposed to model field scattering from metasurfaces are FDTD and FDFD \cite{Caloz_MS_Siijm,CalozFDTD, Vaheb_FDTD_GSTC, Smy_Metasurface_Space_Time,FDTD_Metasurface_Disperive}, FEM \cite{Caloz_FEM}, Spectral Domain method \cite{Caloz_Spectral} and Integral Equation Solvers (IE-GSTCs) based on Boundary Element Methods \cite{stewart_BEM}.
Since volumetric numerical methods (FDFD, \sg{FDTD} and FEM) based on finite domains are too inefficient to model electrically large metasurfaces, surface based methods such as IE-GSTCs become a preferred choice.
\ok{In the surface based approach,} metasurface structures are conveniently represented by equivalent zero-thickness models to reduce computational complexity while preserving their physical scattering characteristics.
The \ok{EM properties of the} zero-thickness sheet can then be described in terms of surface susceptibility tensors $\bar{\bm\chi}$ which, in conjunction with the Generalized Sheet Transition Conditions (GSTCs), \ok{provide} a rigorous field model from which to determine the scattered \ok{surface} fields \sg{anywhere in space}
%from the surface
\cite{KuesterGSTC,IdemenDiscont,GSTC_Holloway}.

\ok{While surface based methods have clear computational complexity advantages}, for very large metasurface problems (or alternatively, very high frequencies) they also become unworkable due to excessive resource requirements.
If concessions \tjs{can be} made on the desired level of accuracy, ray methods based on geometrical optics (GO) and a suitable uniform asymptotic description of diffraction effects, \ok{such as the uniform theory of diffraction (UTD) \cite{pathak_UTD}}, emerge as a promising alternative.
UTD based (or similar) \ok{high-frequency} methods are computationally efficient and present an intuitive picture of the fields scattered from the surface.
Consequently, they have been used in \ok{a wide} variety of \ok{applications} including radio propagation modeling \cite{yun_2015} and analysis of EM structures such as impenetrable gratings and diffusers \cite{yvo_2018}, \ok{conventional} curved surfaces \cite{albani_carluccio_pathak_2011}, and diffraction gratings with slowly varying parameters \cite{borovikov_kinber_1994}. 

Motivated by such \ok{applications, a uniform asymptotic description of physical optics (PO) surface scattering} has recently been applied to metasurfaces \cite{yvo}.
\sg{In \cite{yvo}}, a generic metasurface is modeled as a zero-thickness sheet with prescribed reflection and transmission coefficients.
Assuming \ok{that the metasurface is locally periodic}, these coefficients and the corresponding scattered \ok{surface} fields \ok{are expanded into their spatial Fourier components, which are then} incorporated in the EM field integral equations.
Evaluation of these field integrals \ok{in the} high-frequency limit ($k\rightarrow\infty$, \ok{$k=2\pi/\lambda$ being the wavenumber in free space}) leads to an efficient computation of the scattered fields around the metasurface using contributions from \tjs{a} discrete set of critical points.
While the ray-based numerical framework is computationally efficient, the prescribed reflection and transmission coefficients are \ok{in reality} field-dependent, and \ok{should therefore} be \ok{respecified or otherwise updated every time the incident field direction is} %input field angles are
changed. 

It is
%, however, desirable and advantageous
\ok{more desirable} to define the scattering properties of the metasurface in terms of its intrinsic parameters, such as surface susceptibilities $\bar{\bm\chi}$, from which the fields can be computed on demand, for an arbitrary incidence angle. This work, therefore, represents a continuation of the uniform ray description of physical optics scattering by metasurfaces presented in \cite{yvo}, \tjs{in which the GSTCs are directly incorporated into} the
%UTD
\ok{ray}
formulation, \tjs{and the} metasurface is now described in terms of \tjs{{\em angle-independent}} surface susceptibilities $\bar{\bm\chi}$. This \tjs{approach combines the elegance of the numerical framework, the local periodicity of the metasurface and rigorous GSTC boundary conditions to model the wave transformation capabilities of metasurfaces based on surface susceptibilities.}

The paper is organized as follows: Sec.~II describes the problem statement and the GSTC model of \ok{a generic} metasurface.
Sec.~III then \ok{modifies} the GSTCs
%in terms of
%geometrical optical
\ok{for the case of GO fields}
interacting with a \ok{locally periodic} metasurface \ok{in the high-frequency limit.}
%and how \tjs{they are  modified} when local periodicity is assumed.
This entails \ok{(locally)} expressing the \ok{prescribed} %space-dependent
surface \ok{susceptibility profile}
%of the metasurface
in terms of a Fourier series expansion, \ok{and the scattered field as plane-wave spectra of spatial modes}.
\ok{Equations are provided for numerically solving the scattered surface fields for a given GO incident field, and propagating them to arbitrary detector locations away from the surface.}
%Utilizing the GSTCs, the scattered fields are computed and integrated with the UTD framework.
A simple procedure to convert \ok{an arbitrarily} prescribed
%$\bar{\bm\chi}$
\ok{susceptibility profile} into the desired Fourier series form is described in Sec.~IV.
This procedure \ok{uses a spatially windowed Fourier transform, and is} similar to the conventional \red{use of} Short Time Fourier Transforms (STFT).
Several numerical examples are then presented in Sec.~V to demonstrate the proposed \red{ray-optical GSTC (RO-GSTC)} formulation.
Conclusions are finally provided in Sec.~VI.  
\section{Problem Statement}\label{Sec:ProblemStatement}
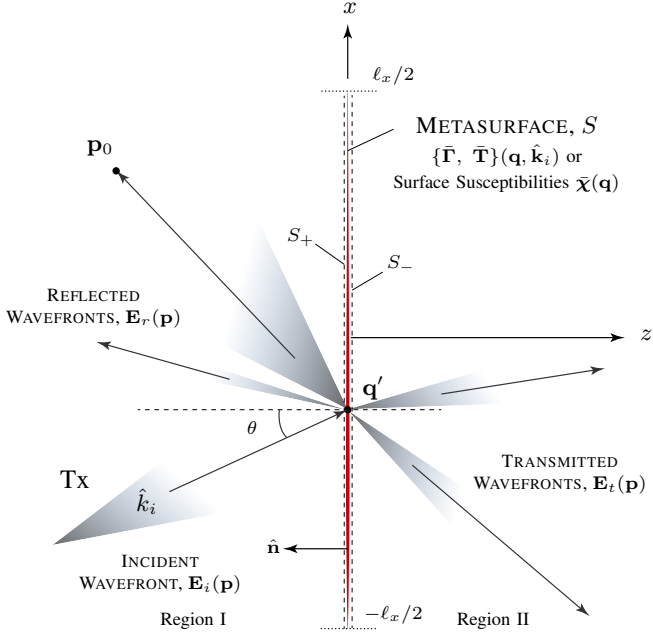
\begin{figure}[!t]
\centering
\begin{overpic}[grid=false, width=\columnwidth]{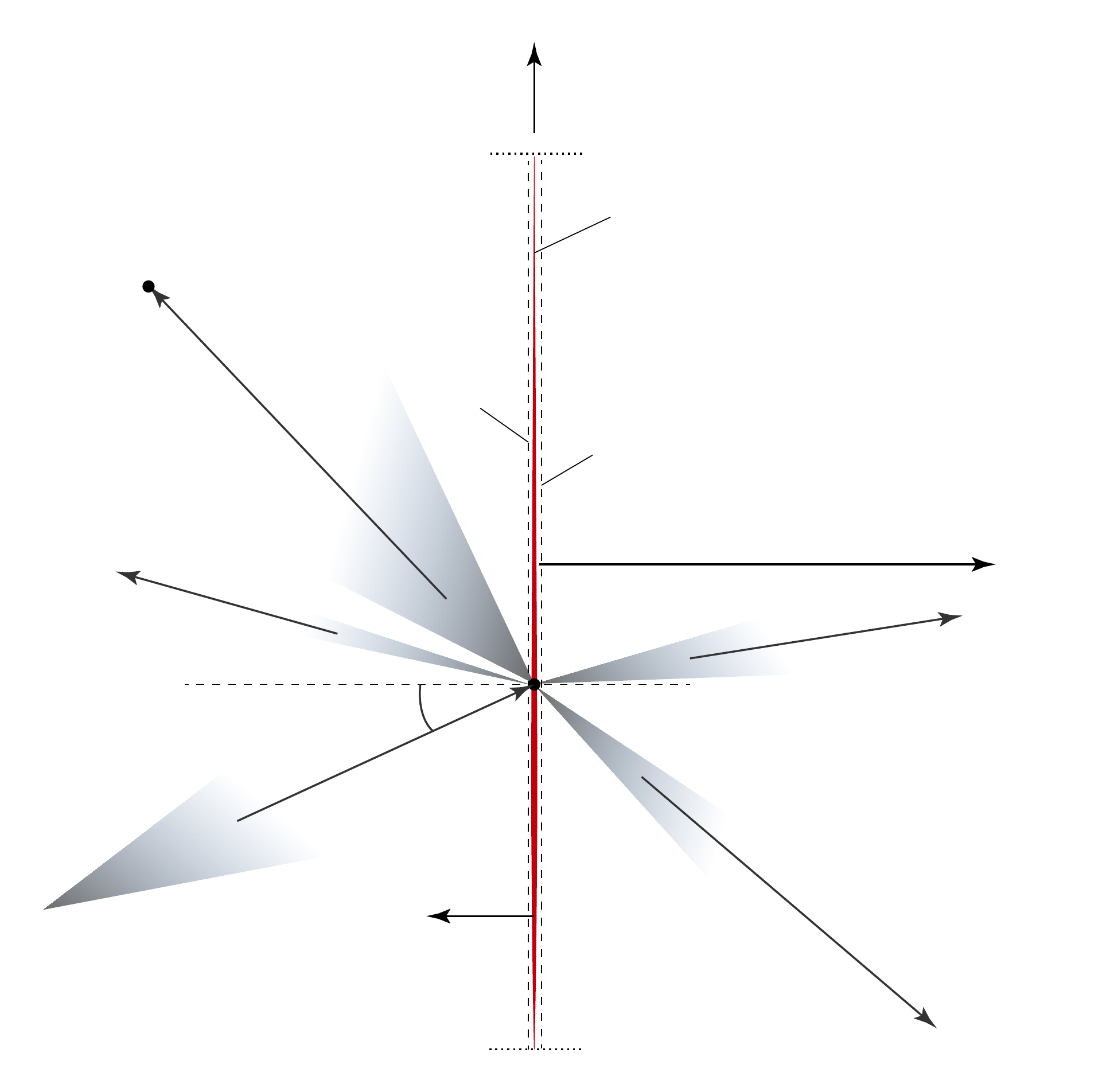}
% \begin{overpic}[grid=false, width=\columnwidth]{Figures/Problem.png}
		\put(72, 75){\makebox(0,0){\small \shortstack{ \textsc{Metasurface, $S$} \\ \scriptsize\ok{$\{\bar{\bm\Gamma},~\bar{\bf T}\}(\q,\hat{\bf k}_i)$} or \\ \scriptsize Surface Susceptibilities \ok{$\bar{\bm\chi}(\q)$}}}}		
		\put(54, 58){\scriptsize $S_-$}
		\put(39,62){\scriptsize $S_+$}
		\put(36,15.25){\scriptsize $\nhat$}
		\put(33,33){\scriptsize $\theta$}
		\put(5, 25){Tx}
		\put(48.5, 97){\makebox(0,0){\small $x$}}
		\put(93, 48){\makebox(0,0){\small $z$}}
		\put(18,23){\makebox(0,0){\small $\hat{k}_i$}}
		\put(52,40){\makebox(0,0){\small $\q'$}}
		\put(11,76){\makebox(0,0){\small $\p_0$}}
		\put(10,52){\makebox(0,0){\scriptsize \textsc{\shortstack{Reflected \\Wavefronts,~$\E_r(\mathbf{p})$}}}}
		\put(20,12){\makebox(0,0){\scriptsize \textsc{\shortstack{Incident \\Wavefront, $\E_i(\mathbf{p})$}}}}
		\put(80,27){\makebox(0,0){\scriptsize \textsc{\shortstack{Transmitted \\Wavefronts,~$\E_t(\mathbf{p})$}}}}
		\put(25,5){\makebox(0,0){\scriptsize Region~I}}
		\put(70,5){\makebox(0,0){\scriptsize Region~II}}
		\put(55,6){\makebox(0,0){\scriptsize $-\ell_x/2$}}
		\put(55,87){\makebox(0,0){\scriptsize $\ell_x/2$}}
\end{overpic} \caption{Diagram showing a non-uniform metasurface interacting with geometrical rays from a source (Tx). The non-uniformity creates multiple scattered rays, whose directions of propagation do not follow Snell's law, that are \sg{received} by different detectors.}
\label{Fig:TxRx2}
\end{figure}

Consider a planar, non-uniform metasurface in the plane spanned by the unit basis vectors $\hat{\bf x}$ and $\hat{\bf y}$, whose normal vector, $\hat{\bf n}$, is \ok{identical to $-\hat{\bf z}$}, as shown in Fig.~\ref{Fig:TxRx2}. A known incident field excites the metasurface, and generates the scattered reflection and transmission fields in free-space Regions I and II, respectively.
The properties of the metasurface are assumed to vary along $\hat{\bf x}$ only \sg{for simplicity}, and the incident field is assumed to be cylindrical and in the $x$--$z$ plane.

The main objective of this paper is to compute the unknown amplitudes of the reflected and transmitted fields at all surface points $\q'$, and then to evaluate \tjs{the} total scattered fields away from the metasurface as a superposition of field contributions originating from the surface points.
The scattering response on the surface can be described in terms of localized transmission and reflection dyadics $\bar{\bm\Gamma}$ and $\bar{\bf T}$ \sg{each point on the surface} \cite{yvo}, well-known to be dependent on the propagation direction of the incident field.
Alternatively, its scattering properties can be described by a well-chosen set of constitutive parameters which are, by definition, field-independent.
The constitutive parameters of choice for metasurfaces are the electric and magnetic surface susceptibility tensors $\bar{\bm\chi}_{\rm ee}$, $\bar{\bm\chi}_{\rm mm}$, $\bar{\bm\chi}_{\rm em}$ and $\bar{\bm\chi}_{\rm me}$, defined by up to 36 complex scalars that capture the microscopic response of the surface to the average fields around it.

For later convenience, we denote the perpendicular and parallel components of an arbitrary vector ${\bf A}$ with respect to the plane with normal $\hat{\bf n}$ by
\begin{gather}
    {\bf A}_{\perp} = A_{\perp} \hat{\bf n} \notag\\
    {\bf A}_{\parallel} = (\bar{\bf I} - \hat{\bf n}\hat{\bf n}) \cdot {\bf A}, \notag
\end{gather}
where $\bar{\bf I}$ is the unit dyadic.
The unit vectors $\hat{\bf u}_1$, $\hat{\bf u}_2$ and $\hat{\bf u}_3$ are defined to be identical to $\hat{\bf x}$, $\hat{\bf y}$ and $\hat{\bf z}$, respectively, and the components of \sg{a vector} ${\bf A}$ along $\hat{\bf u}_p$, $p=\{1,2,3\}$, are denoted by
\begin{equation}
    A_p = ({\bf A})_p = \hat{\bf u}_p \cdot {\bf A}. \notag
\end{equation}
Similarly, the components of a dyadic $\bar{\bf A}$ are denoted by
\begin{equation}
    A_{pq} = (\bar{\bf A})_{pq} = \hat{\bf u}_p \cdot \bar{\bf A} \cdot \hat{\bf u}_q.\notag
\end{equation}
Finally, the superscript $(m)$ denotes the m$^{\text{th}}$ Fourier coefficient of any scalar, vector or dyadic quantity that is a periodic function of $\psi$ with period $\lambda$, i.e.,
\begin{equation}
    \{A, {\bf A}, \bar{\bf A}\}^{(m)} = \frac{1}{\lambda} \int_0^{\lambda} \{A, {\bf A}, \bar{\bf A}\}(\psi) e^{-jkm\psi} d\psi. \label{eq:FourierCoeff}
\end{equation}
%, and an origin, ${\bf q}_0$, in the plane of the surface, such that  with a set of orthonormal basis vectors $\uhat_1$, $\uhat_2$, as:
%
%\begin{align*}
%    \q = \q_0 + u_1\uhat_1 + u_2\uhat_2,
%\end{align*}
%
%\noindent 

\subsection{Field-Dependent $\{\bar{\bm\Gamma}, \bar{\bf T}\}(\q,\hat{\bf k}_i)$}

If the reflection and transmission dyadics are engineered to achieve a desired wave transformation, and are known at each point on the surface for a specified incidence direction $\hat{\bf k}_i$, the GO reflected and transmitted fields can be evaluated as
%
% \begin{subequations}
    \begin{align*}
        \E_r(\q)&=\bar{\GGamma}(\q, \hat{\bf k}_i) \cdot \E_i(\q)\\
        \E_t(\q)&=\bar{\TT}(\q, \hat{\bf k}_i) \cdot \E_i(\q).
    \end{align*}

\noindent In general, $\hat{\bf k}_i$ varies only slowly over the metasurface and may be approximated as being constant over a sufficiently small subdomain, but $\bar{\bm\Gamma}$ and $\bar{\bf T}$ are rapidly varying functions of ${\bf q}$.
If these variations are locally periodic, however, they can be captured by the slowly-varying coefficients of a spatial Fourier expansion.
For $k \rightarrow \infty$, this leads to a convenient representation of the overall scattered field as a collection of ray-optical fields with different amplitudes, propagation directions and radii of curvature \cite{yvo}.
These rays can then be propagated in free space away from the metasurface using standard ray theory, to construct the total scattered fields anywhere in space.

% tjs -- I think at this point we don't need to bring up this. It is only an issue when we use the SFFT method.
% \red{\textbf{[Should they be not only constant, but their phase variation over distance be negligible? See Sec. IV.]}}. \scott{\textbf{[They can be approximated as constant with their phase variation over the short distance being negligible.]}} 
%
%Consequently, the scattered fields also take the similar form, which in turn provides a convenient representation of the outgoing wavefronts as a collection of several optical rays with different directions, amplitudes and radii of curvatures. These rays then can be propagated in free-space away from the metasurface using standard UTD principles, to construct the total scattered fields anywhere in space.

\tjs{This} representation of the metasurface in terms of dyadic transmission and reflection coefficients leads to a numerically convenient solution compatible with standard ray-tracing techniques. \tjs{However,} when attempting to model a system of surfaces with a large range of incidence angles (such as when other scattering objects are present or for a more complicated incident field), it is disadvantageous having to store the scattering coefficients as functions of $\hat{\bf k}_i$ (or equivalently, as functions of the local spherical coordinates $\theta$ and $\phi$) for all points on the surface. Such a representation introduces both numerical error and large memory requirements.
A more elegant representation of such metasurfaces is to go one step deeper \tjs{into the physical description} and represent them using their angle-independent constitutive parameters, i.e. surface susceptibilities $\bar{\bm\chi}(\q)$. \tjs{This has an additional advantage of providing} a more intuitive description of the surface interaction in terms of its microscopic dipolar responses. 
\subsection{Field-Independent Surface Susceptibilities, $\bar{\bm\chi}(\q)$}
Metasurfaces represent a spatial discontinuity in both the phase and magnitude of a wave. As metasurfaces have a sub-wavelength thickness, they can be modeled as a zero-thickness surface using the Generalized Sheet Transition Conditions (GSTCs) \cite{IdemenDiscont}\cite{KuesterGSTC}. The GSTCs relate the differential and average fields across the metasurface via the following relations:
\begin{subequations}\label{Eq:GSTCs}
\begin{align}
    \nhat \times \Delta \H &= j\omega\P_{\parallel} - \nhat \times \nabla_{\parallel}M_{\perp}\label{Eq:GSTCDeltaH} \\
    \Delta \E \times \nhat &= j\omega\mu\M_{\parallel} - \nabla_{\parallel}\left(\frac{P_{\perp}}{\epsilon}\right)\times \nhat\label{Eq:GSTCDeltaE},
\end{align}
\end{subequations}
\noindent where
%$\Delta\Psi$
\begin{equation}
    (\Delta{\bf E}, \Delta{\bf H}) = (
    {\bf E}^{+} - {\bf E}^{-},
    {\bf H}^{+} - {\bf H}^{-})\notag
\end{equation}
are the differences between the electric and magnetic fields on the positive (illuminated) and negative (shadowed) sides of the metasurface, respectively.
Furthermore, $\epsilon$ and $\mu$ are the electric permittivity and magnetic permeability of free space, respectively, and $\omega$ denotes the angular frequency.
The operator $\nabla_{||}(\cdot)$ returns the gradient of its argument in the plane of the surface.
$\P$ and $\M$ are the electric and magnetic surface polarization densities, given by
\begin{subequations}\label{Eq:PMs}
\begin{align}
    \P &= \epsilon\bar{\bm\chi}_{\rm ee} \cdot \E_\text{av}
    +\sqrt{\epsilon\mu}\bar{\bm\chi}_{\rm em} \cdot \H_\text{av}\\
    \M &= \bar{\bm\chi}_{\rm mm} \cdot \H_\text{av} + \sqrt{\frac{\epsilon}{\mu}}\bar{\bm\chi}_{\rm me} \cdot \E_\text{av},
\end{align}
\end{subequations}
\noindent
\noindent where
%$\Delta\Psi$
\begin{equation}
    ({\bf E}_{\rm av}, {\bf H}_{\rm av}) = \frac{1}{2}(
    {\bf E}^{+} + {\bf E}^{-},
    {\bf H}^{+} + {\bf H}^{-})\notag
\end{equation}
are the average fields across the surface.

The dyadics $\bar{\bm\chi}_{ab}$, with $a,b\in\{\text{e,m}\}$, represent the fundamental constitutive parameters of an arbitrarily specified metasurface (linear and time-invariant), and \tjs{are} thus \emph{independent} of the input fields (and the angle of incidence) \cite{metasurface_synthesis}. For a given set of $\bar{\bm\chi}_{ab}$ and the specified incident fields, $\E_i(\q)$, the scattered fields (and if desired, $\mathbf{\Gamma}$ and $\mathbf{T}$) may be rigorously obtained by solving \eqref{Eq:GSTCs} using \eqref{Eq:PMs} \cite{stewart_smy_gupta_2018,stewart_BEM}. This angle independent property of \tjs{the} surface susceptibilities \tjs{therefore} enables modeling an arbitrary metasurface using compact matrices, \tjs{which can subsequently be integrated into} a ray-based simulation environment. It should be noted that for purely tangential surface polarizabilities, a mathematically equivalent formulation in terms of location-dependent electric and magnetic sheet impedances \cite{Chi_Review} may alternatively be used, as in \cite[Sec.~IV]{yvo}. Surface susceptibilities on the other hand, represent a completely general surface characterization, including both tangential and normal polarization in its complete form \cite{KuesterGSTC, Chi_Review}  \sg{necessary to correctly obtain the angular response of a metasurfaces} -- hence is the preferred choice here.

\section{Asymptotic Solution of Locally Periodic GSTCs}\label{Sec:AsymptoticSolution}

\subsection{Local Periodicity}
In preparation of using surface susceptibilities in a high-frequency (ray-optical) description of the scattered fields,
%UTD based ray tracing environment,
consider again the problem setup of Fig.~\ref{Fig:TxRx2}, but now assume that the properties of the metasurface exhibit sub-wavelength location-dependent periodicity \cite{yvo} along $\hat{\bf x}$.
Under this assumption, the susceptibility $\bar{\bm\chi}_{ab}$ may locally be expressed as a periodic function of a location-dependent phase parameter $\psi$.
It can therefore be expanded in a Fourier series \sg{as}
\begin{align}\label{Eq:ChiFourierSeries}
    \ok{\bar{\bm\chi}_{ab}(x)}\; {\approx}\;
    \bar{\bm\chi}_{ab}(\psi;x') = \sum_{m=-\infty}^{\infty}\bar{\bm\chi}_{ab}^{(m)}(x')e^{jkm\psi(x)}
\end{align}
in any sufficiently small subdomain of $S$ around $x'$,
where $m$ is an integer index referred to as the spatial mode number.
${\bm\chi}_{ab}$ has a constant period of $\lambda$ with respect to $\psi$, but its local period with respect to $x$ is determined by the function $\psi(x)$, specifically its slope at $x'$.
Assuming $\psi(x)$ to be continuous and smooth, it can be expanded around $x'$ as
\begin{equation}\label{Eq:PsiQuadratic}
    \psi(x) \simeq \psi(x') + \dot{\psi}(x')(x-x') + \frac{1}{2}\ddot{\psi}(x')(x-x')^2,
\end{equation}
where $\dot{\psi}$ and $\ddot{\psi}$ denote its first and second spatial derivatives.
In addition to its dependence on $\psi$, ${\bm\chi}_{ab}$ may also vary (slowly) as a function of $x'$, allowing further non-uniformity of its scattering properties.
In practice, the infinite sum in (\ref{Eq:ChiFourierSeries}) can be truncated so that $|m| \leq M$, where $M$ is a mode truncation number that must be chosen large enough to avoid significant loss of accuracy.
These summation limits are suppressed in the remainder of this paper.
%
%
%If the metasurface is \emph{locally periodic},  following \tjs{a similar methodology as done with the transmission and reflection coefficients, $\mathbf{T}$ and $\mathbf{\Gamma}$ (see \eqref{Eq:TRFourier}}) the surface susceptibility $\bar{\chi}_{ab}$ may be expressed in terms of Fourier series expansion as 
%
%\begin{align}\label{Eq:ChiFourierSeries}
%    \bar{\chi}_{ab}(\psi;\q') = %\sum_{m=-\infty}^{\infty}\bar{\chi}_{ab}^{(m)}(\q')e^{jkm\psi},
%\end{align}
%
%\noindent with the corresponding Fourier coefficients, $\bar{\chi}_{ab}^{(m)}(\q')$ given by,
%
%\begin{align}\label{Eq:ChiFourierCoefficients}
%    \bar{\chi}_{ab}^{(m)}(\q') = \frac{1}{\lambda}\int_0^\lambda %\bar{\chi}_{ab}(\psi;\q')e^{-jkm\psi}d\psi,
%\end{align}
%
%\noindent in terms of the parametric variable $\psi(\q)$.
%

For a special case of a periodic metasurface with a uniform period $\Lambda$ with respect to $x$, such that $x/\Lambda = \psi/\lambda$, \eqref{Eq:ChiFourierSeries} reduces to its standard form \cite{tiukuvaara_2020, Ville_Floquet}, with
\begin{align*}
k\psi = k\frac{\lambda x}{\Lambda} = \frac{2 \pi}{\Lambda} x.
\end{align*}

%
%To facilitate a seamless integration of the surface susceptibility description of the surface in the UTD formulation, let us assume that it is possible to approximate and represent an arbitrarily specified surface susceptibility tensor $\bar{\chi}_{ab}(\q)$ of the metasurface in the form of \eqref{Eq:ChiFourierSeries}.
%*** This means that the surface susceptibility tensors are locally periodic functions (with a period of $\lambda$) of the phase function $\psi$, following \eqref{Eq:ChiFourierSeries}.
%Note that, consistent with the problem statement, $\psi$ depends only on the surface coordinate $x$, and this dependence varies slowly with respect to $x'$. ***
A detailed procedure to estimate the parameters of (\ref{Eq:ChiFourierSeries}), i.e., the function $\psi(x)$ and the modal susceptibilities $\bar{\bm\chi}^{(m)}_{ab}$, $a,b\in\{{\rm e}, {\rm m}\}$, from a given susceptibility profile is described and discussed in Sec.~IV.
%
%\subsection{GSTCs for UTD Integration}\label{Sec:MetasurfaceIntegration}
\subsection{High-Frequency Approximation}\label{Sec:MetasurfaceIntegration}

We would like to expand the transmitted and reflected fields due to the metasurface into plane-wave spatial harmonics according to Floquet's theorem for periodic surfaces ~\cite{tiukuvaara_2020} \cite{ishimaru_2017}.
Strictly speaking, the results of this theorem are not applicable to the problem addressed herein, not only because (\ref{Eq:ChiFourierSeries}) allows the metasurface periodicity to be non-uniform, but also because incident and scattered fields are not restricted to being planar, and edge (diffraction) effects are taken into account.
If the metasurface is electrically large, however, the incident and scattered fields can be well approximated by their asymptotic behavior in the high-frequency limit, giving rise to a ray-optical field representation governed by geometrical optics (GO) and the geometrical theory of diffraction (GTD) or any of its uniform extensions \cite{balanis_2012,yvo}.

In the high-frequency limit ($k\rightarrow\infty$), wave interaction is a local phenomenon, and depends only on the surface and field properties in an infinitesimally small area around the incidence point \cite{balanis_2012}.
The incident electric surface field in the vicinity of a reference point $x'$ may be written as
\begin{equation}
\E_i(x) = \E_i(x') e^{-jk\varphi_i(x)},
\end{equation}
where
\begin{equation}\label{Eq:Phii}
    \varphi_i \simeq (\hat{\bf k}_i \cdot \hat{\bf x})x
    + \frac{(x-x')^2}{2\rho_i}
\end{equation}
is a phase function that contains the radius of curvature, $\rho_i$, describing the divergence (or convergence) of the incident wavefront at $x'$.
The scattered surface fields may be locally approximated as plane-wave spectra
\begin{subequations}\label{Eq:ScatteredFieldEquations}
    \begin{align}
        \E_r(x) &= \sum_m \E_r^{(m)}(x')e^{jkm\psi(x)}\\
        \E_t(x) &= \sum_m \E_t^{(m)}(x')e^{jkm\psi(x)},
    \end{align}
\end{subequations}
\noindent and the corresponding magnetic fields are obtained via the local plane-wave approximation
\begin{gather}
    \H_i(x') = \frac{1}{\eta} \hat{\bf k}_i \times \E_i(x') \\
    \label{Eq:ScatteringDirection}
    \H_a^{(m)}(x')=\frac{1}{\eta}\khat_a^{(m)}\times\E_a^{(m)}(x'),
\end{gather}
where $a\in\{r,t\}$ and $\eta=\sqrt{\mu/\epsilon}$ is the free-space wave impedance.
Furthermore,
$\khat_a^{(m)}$ denotes the (real-valued) propagation direction vectors corresponding to $({\bf E}_a^{(m)}, {\bf H}_a^{(m)})$, related to $\hat{\bf k}_i$ by
\begin{align}\label{eq:proj2}
    \bigl( \hat{\bf k}_a^{(m)} \bigr)_{\parallel} =
    \bigl( \hat{\bf k}_i \bigr)_{\parallel} - m \dot{\psi} \, \hat{\bf x},
\end{align}
provided that the magnitude of the right-hand side of (\ref{eq:proj2}) does not exceed one; modes for which this is the case are evanescent and propagate parallel to the surface \cite{yvo}. \sg{This is essentially the phase matching condition applied locally to the surface.}

The surface fields ${\bf E}^{\pm}$ and ${\bf H}^{\pm}$ can now be written as
\begin{subequations}\label{Eq:EpmHpm}
\begin{align}
({\bf E}^{+}, {\bf H}^{+}) &= ({\bf E}_i, {\bf H}_i) + \sum_m ({\bf E}_r^{(m)}, {\bf H}_r^{(m)}) e^{jkm\psi} \label{eq:Er}\\
({\bf E}^{-}, {\bf H}^{-}) &= \sum_m ({\bf E}_t^{(m)}, {\bf H}_t^{(m)}) e^{jkm\psi}. \label{eq:Et}
\end{align}
\end{subequations}

\subsection{Solution of GSTC Surface Field Equations}\label{Eq:GSTC_Solution}
\label{sec:IIIC}
Substituting the locally periodic surface fields \eqref{Eq:EpmHpm} and susceptibilities \eqref{Eq:ChiFourierSeries}, the GSTCs \eqref{Eq:GSTCs} can be rewritten as
%
%\red{\textbf{All the following equations must be checked and verified.}}
%
%\begin{subequations}\label{Eq:GSTCs_EHpm}
%\begin{align}
%    \sum_{q=1}^3
%    \Bigl[
%    &A_{pq}(\hat{\bf k}_i)I_q \nonumber\\
%    &+ \sum_m\Bigl(
%    A_{pq}(\hat{\bf k}_r^{(m)}) \Gamma_q^{(m)} -
%    A_{pq}(\hat{\bf k}_t^{(m)}) T_q^{(m)}
%    \Bigr) e^{jkm\psi}
%    \Bigr] \nonumber\\
%    &= \frac{jk}{\epsilon} P_p - \eta \delta_{n-2} M'_3 \\
%%
%    \sum_{q=1}^3 &B_{pq}
%    \Bigl[
%    I_q + \sum_m\Bigl(
%    \Gamma_q^{(m)} - T_q^{(m)}
%    \Bigr) e^{jkm\psi}
%    \Bigr] \nonumber\\
%    &= jk\eta M_p + \frac{1}{\epsilon} \delta_{n-2} P'_3
%\end{align}
%\end{subequations}
%for $p=\{1,2\}$, in which
%\begin{subequations}
%\begin{align}
%    \Gamma_q^{(m)} &= \hat{\bf u}_q \cdot \bar{\bm\Gamma}_q^{(m)} \\
%    T_q^{(m)} &= \hat{\bf u}_q \cdot \bar{\bf T}_q^{(m)}
%\end{align}
%\end{subequations}
%are the unknown variables.
%
\begin{subequations}\label{Eq:GSTCs_EHpm}
\begin{align}
    &\Big[\bar{\A}(\khat_i)\tjs{\cdot\E_i}\nonumber\\
    &\quad+\sum_m\left(\bar{\A}(\khat_r^{(m)})\cdot \E_r^{(m)} - \bar{\A}(\khat_t^{(m})\cdot\E_t^{(m)}\right) e^{jkm\psi}\Big]\nonumber\\
    &\quad= \frac{jk}{\epsilon} \P_{\parallel}
    -\eta \, \hat{\bf n} \times \nabla_{\parallel}(\hat{\bf n} \cdot \M) \\[1em]
    &\bar{\bf B} \cdot \Big[
    \tjs{\E_i} + \sum_m\left(\E_r^{(m)}-\E_t^{(m)}\right)
    e^{jkm\psi} \Big] \nonumber\\
    &\quad= jk\eta \M_{\parallel}
    + \frac{1}{\epsilon} \hat{\bf n} \times \nabla_{\parallel}(\hat{\bf n} \cdot \P),
%    -\nhat&\times\Big[
%    \tjs{\E_i} + \sum_m\left(\E_{rm}-\E_{tm}\right)
%    e^{jkm\psi} \Big] \nonumber\\
%    &= jk\eta \M_{\parallel}
%    + \frac{1}{\epsilon} \hat{\bf n} \times \nabla_{\parallel}(\hat{\bf n} %\cdot \P),
\end{align}
\end{subequations}
\noindent in which
\begin{align*}
    \P &= \epsilon\Big[\bar{\X}_e(\khat_i)\tjs{\cdot\E_i}\\
    &+\sum_m\left(\bar{\X}_e(\khat_r^{(m)}) \cdot \E_r^{(m)} + \bar{\X}_e(\khat_t^{(m)}) \cdot \E_t^{(m)} \right) e^{jkm\psi}\Big]\nonumber\\[1em]
%\end{align*}
%
%\begin{align*}
    \M &= \frac{1}{\eta}\Big[\bar{\X}_e(\khat_i)\tjs{\cdot\E_i}\\
    &+\sum_m\left(\bar{\X}_m(\khat_r^{(m)}) \cdot \E_r^{(m)} + \bar{\X}_m(\khat_t^{(m)}) \cdot \E_t^{(m)} \right) e^{jkm\psi}\Big]\nonumber
\end{align*}
\noindent with 
\begin{gather*}
    \bar{\A}(\khat) = \khat\nhat-(\khat\cdot\nhat)\barI \\
    \bar{\bf B} = -\hat{\bf n} \times \bar{\bf I}
\end{gather*}
\noindent and
\begin{align*}
    \bar{\X}_{\rm e}(\khat) &= \frac{1}{2} \bigl(
    \bar{\bm\chi}_{\rm ee} + \bar{\bm\chi}_{\rm em} \times \khat
    \bigr)\\
    \bar{\X}_{\rm m}(\khat) &= \frac{1}{2} \bigl(
    \bar{\bm\chi}_{\rm mm} \times \khat + \bar{\bm\chi}_{\rm me}
    \bigr).
\end{align*}

The vector equations \eqref{Eq:GSTCs_EHpm} can be turned into an equivalent set of (four) scalar equations by pre-multiplying both by $\hat{\bf u}_p$ with $p=\{1,2\}$.
Because these equations are periodic in $\psi$, they can be more conveniently solved in the spatial Fourier domain, i.e., after
multiplying each by $e^{-jkn\psi}$ and integrating, with respect to $\psi$, from 0 to $\lambda$.
This yields
\begin{subequations}\label{Eq:LinearFieldEquations}
\begin{align}
\sum_{q=1}^3
&\Bigl[
A_{pq}(\hat{\bf k}_i) \delta_n E_{i,q}
+ 
A_{pq}(\hat{\bf k}_r^{(n)}) \tjs{E^{(n)}_{r,q}} - 
A_{pq}(\hat{\bf k}_t^{(n)}) \tjs{E^{(n)}_{t,q}}
\Bigr] \nonumber\\
&=
\frac{jk}{\epsilon} P_p^{(n)}
% j\omega \, \hat{\bf u}_p \cdot {\bf P}^{(n)}
- \eta \delta_{p-2} \dot{M}^{(n)}_{\perp}
\label{eq:gstc_scalar1} \\
\sum_{q=1}^3 &B_{pq} \left[
\delta_n E_{i,q} + \tjs{E^{(n)}_{r,q}} - \tjs{E^{(n)}_{t,q}}
\right]
=
jk\eta M_p^{(n)}
%j\omega\mu (\hat{\bf u}_p \times \hat{\bf n}) \cdot {\bf M}^{(n)}
+\frac{1}{\epsilon} \delta_{p-2} \dot{P}^{(n)}_{\perp},
\label{eq:gstc_scalar2}
\end{align}
\end{subequations}
in which $\delta_n$ denotes the Kronecker delta function, and
\begin{subequations}\label{Eq:LinearEq2}
\begin{align}
&P_p^{(n)}
= \epsilon \sum_{q=1}^3 \Bigl[
X_{{\rm e}, pq}^{(n)}(\hat{\bf k}_i) E_{i,q} \\
&+ \sum_m
\left(
X_{{\rm e}, pq}^{(n-m)}(\hat{\bf k}_r^{(m)}) E^{(m)}_{r,q} +
X_{{\rm e}, pq}^{(n-m)}(\hat{\bf k}_t^{(m)}) E^{(m)}_{t,q}
\right)
\Bigr]
\label{eq:gstc_scalar3} \nonumber\\
&M_p^{(n)}
= \frac{1}{\eta} \sum_{q=1}^3 \Bigl[
X_{{\rm m},pq}^{(n)}(\hat{\bf k}_i) E_{i,q} \\
&+ \sum_m
\left(
X_{{\rm m},pq}^{(n-m)}(\hat{\bf k}_r^{(m)}) \tjs{E^{(m)}_{r,q}} +
X_{{\rm m},pq}^{(n-m)}(\hat{\bf k}_t^{(m)}) \tjs{E^{(m)}_{t,q}}
\right)
\Bigr]
\label{eq:gstc_scalar4} \nonumber\\
&\dot{P}^{(n)}_{\perp}
= -jk\epsilon \sum_{q=1}^3 
\biggl[
k_{\parallel}^{(0)} X_{{\rm e},3q}^{(n)}(\hat{\bf k}_i) E_{i,q}
+ \sum_m
k_{\parallel}^{(m)} \nonumber\\
&\times\left(
X_{{\rm e},3q}^{(n-m)}(\hat{\bf k}_r^{(m)}) \tjs{E_{rm,q}} +
X_{{\rm e},3q}^{(n-m)}(\hat{\bf k}_t^{(m)}) \tjs{E_{tm,q}}
\right)
\biggr]
 \\
%\hspace{-3mm}
%
&\dot{M}^{(n)}_{\perp}
= -\frac{jk}{\eta} \sum_{q=1}^3
\biggl[
k_{\parallel}^{(0)} X_{{\rm m},3q}^{(n)}(\hat{\bf k}_i) E_{i,q}
+ \sum_m
k_{\parallel}^{(m)} \nonumber\\
&\times\left(
X_{{\rm m},3q}^{(n-m)}(\hat{\bf k}_r^{(m)}) \tjs{E_{rm,q}} +
X_{{\rm m},3q}^{(n-m)}(\hat{\bf k}_t^{(m)}) \tjs{E_{tm,q}}
\right)
\biggr]. 
\end{align}
\end{subequations}
In (\ref{Eq:LinearFieldEquations}) and (\ref{Eq:LinearEq2}),
\begin{gather*}
E^{(m)}_{r,q} = (\E^{(m)}_r)_q \\
E^{(m)}_{t,q} = (\E^{(m)}_t)_q
\end{gather*}
are the unknown variables of interest, and
\begin{gather*}
E_{i,q} = ({\bf E}_i)_q \\
X^{(n)}_{{\rm e},pq}(\hat{\bf k}) = (\bar{\bf X}_{\rm e}(\hat{\bf k}))_{pq}^{(n)} \\
X^{(n)}_{{\rm m},pq}(\hat{\bf k}) = (\bar{\bf X}_{\rm m}(\hat{\bf k}))_{pq}^{(n)}
\end{gather*}
and
%
%\begin{align*}
%E_{i,q} = ({\bf E}_i)_q
%\end{align*}
\begin{align*}
k_{\parallel}^{(m)} = \hat{\bf k}_i \cdot \hat{\bf x} - m \dot{\psi}
\end{align*}
%\begin{align*}
%A_{pq}(\hat{\bf k}) =
%\hat{\bf u}_p \cdot \bar{\bf A}(\hat{\bf k}) \cdot \hat{\bf u}_q
%\end{align*}
%\begin{align*}
%B_{pq} = (\hat{\bf u}_p \times \hat{\bf u}_q) \cdot \hat{\bf u}_3
%\end{align*}
%\begin{align*}
%X^{(n)}_{{\rm e},pq}(\hat{\bf k}) = (\bar{\bf X}_{\rm e}(\hat{\bf %k}))_{pq}^{(n)}
%\hat{\bf u}_p \cdot \bar{\bf X}^{(n)}_{\rm e}(\hat{\bf k}) \cdot \hat{\bf u}_q
%\frac{1}{2} \chi_{ee,np} +
%\frac{1}{2} \sum_{q=1}^3 \chi_{em,nq} (\hat{\bf u}_p \times \hat{\bf u}_q) \cdot \hat{\bf k}
%\end{align*}
%\begin{align*}
%X^{(n)}_{{\rm m},pq}(\hat{\bf k}) = (\bar{\bf X}_{\rm m}(\hat{\bf k}))_{pq}^{(n)}
%\hat{\bf u}_p \cdot \bar{\bf X}^{(n)}_{\rm m}(\hat{\bf k}) \cdot \hat{\bf u}_q
%\frac{1}{2} \chi_{me,np} +
%\frac{1}{2} \sum_{q=1}^3 \chi_{mm,nq} (\hat{\bf u}_p \times \hat{\bf u}_q) \cdot \hat{\bf k}
%\end{align*}
%\begin{align*}
%\bar{\bf X}^{(n)}_{\rm e}(\hat{\bf k}) =
%\frac{1}{2} \int_0^{\lambda} \left[
%\bar{\bm\chi}_{\rm ee}(\psi) + \bar{\bm\chi}_{\rm em}(\psi) \times \hat{\bf k}
%\right] e^{-jkn\psi} d\psi
%\end{align*}
%\begin{align*}
%\bar{\bf X}^{(n)}_{\rm m}(\hat{\bf k}) =
%\frac{1}{2} \int_0^{\lambda} \left[
%\bar{\bm\chi}_{\rm mm}(\psi) \times \hat{\bf k} + \bar{\bm\chi}_{\rm me}(\psi)
%\right] e^{-jkn\psi} d\psi
%\end{align*}
can be computed {\it a priori}.
Substituting (\ref{Eq:LinearEq2}) into \eqref{Eq:LinearFieldEquations}
%for \mbox{$p=\{1,2\}$} and $-M \leq m \leq M$
results in a system of
% $4(2M+1)$
linear equations which can be numerically solved for $E^{(m)}_{r,q}$ and $E^{(m)}_{t,q}$, $q=\{1,2,3\}$, $|m| \leq M$, by also requiring that
\begin{equation}\label{Eq:TEM}
\sum_{q=1}^3 (\hat{\bf k}_r^{(m)})_q \tjs{E^{(m)}_{r,q}} = 
\sum_{q=1}^3 (\hat{\bf k}_t^{(m)})_q \tjs{E^{(m)}_{t,q}} = 0.
\end{equation}
Eq.~(\ref{Eq:TEM}) follows from the fact that each reflected and transmitted plane-wave electric field component must be perpendicular to its corresponding propagation direction.
This completes the determination of the unknown ray-optical scattered surface fields, which can now be propagated to the far field \sg{anywhere in space}. 

\subsection{Forward Ray Tracing to Far-Field Detector}
In order to determine the total field received at an arbitrary detector location $(x,z)$ away from the metasurface $S$ (i.e., $z \neq 0$), the Huygens contributions from $S$ must be integrated and added to the incident field (in the absence of $S$).
An expression for the radiative scattered field is given by the Stratton-Chu integral equation \cite{RothWell}
\begin{align}
\E_s = jk\iint_S \bigl[
\hat{\r} \times \M + \eta\,
\hat{\r}\times\hat{\r} \times \J
\bigl]
\frac{e^{-jkr}}{4\pi r}dS, \label{Eq:Stratton}
\end{align}
where ${\bf r}=r\hat{\bf r}$ is the vector from the integration point to $(x,z)$, and
\begin{gather}
    {\bf M} = (\E^{+} - \E^{-}) \times \hat{\bf n}\nonumber\\
    {\bf J} = \hat{\bf n} \times (\H^{+} - \H^{-})\nonumber
\end{gather}
are the net equivalent magnetic and electric currents on $S$, which can be evaluated by substituting the reflected and transmitted surface fields determined in the previous subsection into (\ref{Eq:EpmHpm}).
%
%
%where the equivalent magnetic and electric current on the surface $S$ are $\M^\pm=\mp(\hat{\mathbf{n}}\times \E^\pm)$, $\J^\pm= \pm(\hat{\mathbf{n}}\times \H^\pm)$, respectively in terms of fields on each side of the surface, \tjs{with} $S_+$ and $S_-$ as shown in Fig.~\ref{Fig:TxRx2}, and $\r = r~\hat{\r}$ is the vector from $\q_c$ to $\p$. Under the Physical Optics (PO) Approximation, the surface fields are approximated as $\{\E_i,\mathbf{H}_i\} + \{\E_r,\mathbf{H}_r\}$ on $S_+$ and $\{\E_t,\mathbf{H}_t\}$ on $S_-$, which are of the forms \tjs{given in \eqref{Eq:ScatteredFieldEquations}, \eqref{Eq:ScatteringDirection} and \eqref{Eq:EpmHpm}}
%\red{\textbf{[Why is it an approximation? Due to finite edges and thus edge diffraction? Due to ray optical description?]}}. 
%Using \eqref{Eq:Stratton} with these PO fields, the scattered fields $\E_s(\p)$ must now be calculated at the observation point $\p$.
Eq.~(\ref{Eq:Stratton}) can then be rewritten as a sum \cite{yvo} 
\begin{equation}
    \E_s = \E_{s,i}^{(0)} + \sum_m \E_{s,r}^{(m)} + \sum_m \E_{s,t}^{(m)}
\end{equation}
of distinct scattering components, with
\begin{align}
    \E_{s,i}^{(0)} &= +jk \iint_S \bar{\bf G}(\hat{\bf k}_i, \hat{\bf r}) \cdot \E_i \frac{e^{-jkr}}{4\pi r} dS \nonumber\\
%\end{equation}
%\begin{equation}
    \E_{s,r}^{(m)} &= +jk \iint_S \bar{\bf G}(\hat{\bf k}^{(m)}_r, \hat{\bf r}) \cdot \E_r^{(m)} e^{jkm\psi} \frac{e^{-jkr}}{4\pi r} dS \nonumber\\
%\end{equation}
%\begin{equation}
    \E_{s,t}^{(m)} &= -jk \iint_S \bar{\bf G}(\hat{\bf k}^{(m)}_t, \hat{\bf r}) \cdot \E_t^{(m)} e^{jkm\psi} \frac{e^{-jkr}}{4\pi r} dS \nonumber
\end{align}
and \cite{albani_carluccio_pathak_2011}
\begin{align*}
\bar{\G}(\mathbf{\hat k},\hat{\bf r}) = (\nhat \cdot \rhat)\barI - \nhat\rhat + (\barI -\rhat\rhat)\cdot\left[({\hat{\bf k}} \cdot \nhat)\barI - \hat{\bf k}\nhat \right].
\end{align*}

\begin{figure}[t]
\centering
\begin{overpic}[grid=false, scale = 0.65]{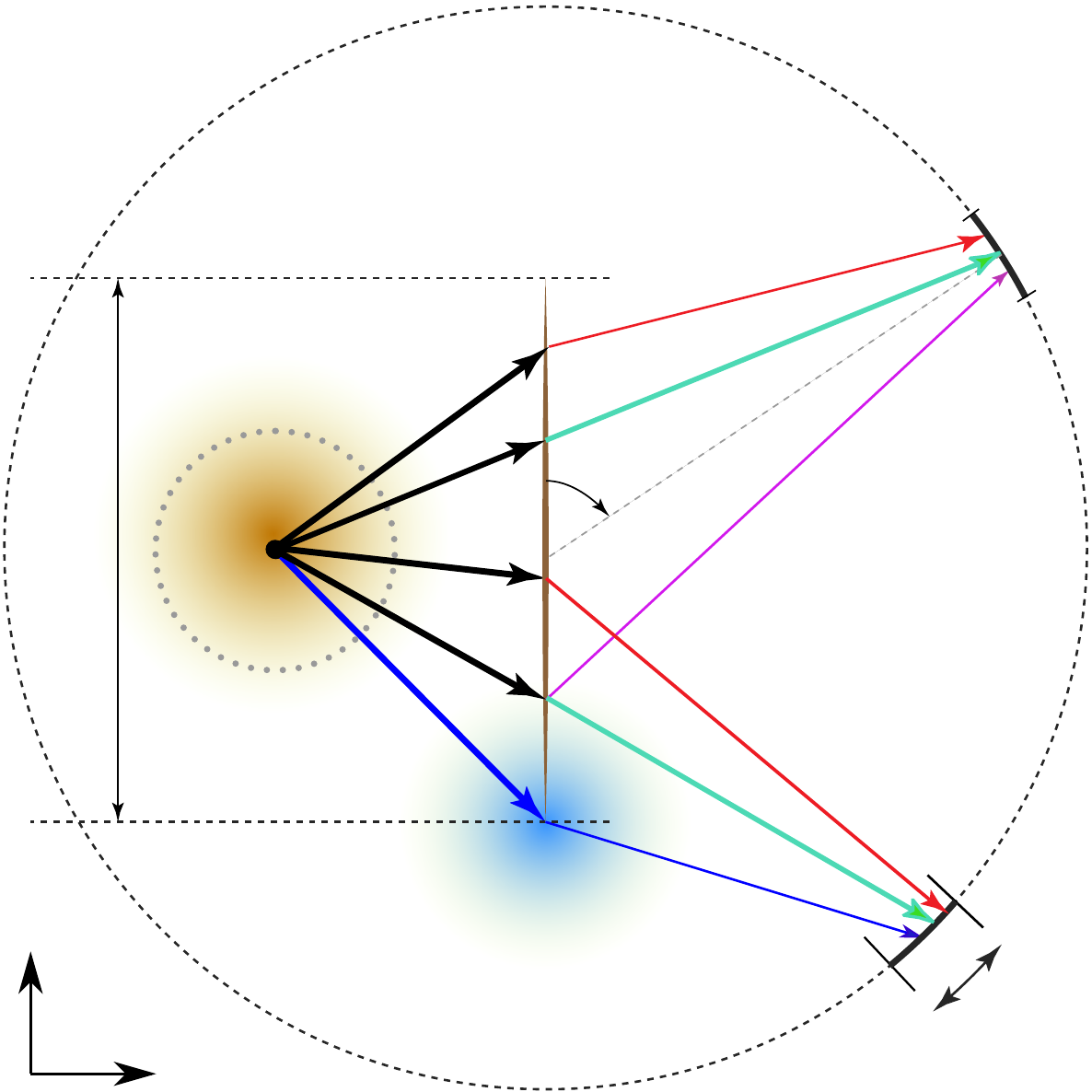}
% \begin{overpic}[grid=false, scale = 0.6]{Figures/FRTSimSetup3.png}
%		\put(80, 40){\scriptsize \rotatebox{270}{$L_y$}}
%		\put(56, 59){\scriptsize \rotatebox{50}{$\shat_t^{(1)}$}}
%		\put(35, 66){\scriptsize \rotatebox{-50}{$\shat_t^{(-1)}$}}
%		\put(51, 66){\scriptsize $\shat_t^{(0)}$}
    		\put(15, 1){\scriptsize $z$}
    		\put(1, 14){\scriptsize $x$}
%		\put(43,46){\scriptsize $L_x$}
		\put(40, 15){\scriptsize \shortstack{Edge\\ Diffraction}}

		\put(5, 46){\scriptsize $\ell_x$}
		\put(93, 78){\scriptsize D~\#1}
		\put(92, 16){\scriptsize D~\#2}
		\put(88, 6){\scriptsize \rotatebox{45}{$\Delta\phi$}}
		\put(64, 55){\scriptsize \rotatebox{0}{$R$}}
		\put(53, 57){\scriptsize $\phi$}
		\put(42, 78){\scriptsize S~$(z=0)$}
		\put(35, 34){\scriptsize $r_0$}
		\put(39, 38){\scriptsize $r_1$}
		\put(40, 45){\scriptsize $r_2$}
		\put(40, 53){\scriptsize $r_3$}
		\put(37, 63){\scriptsize $r_4$}
		\put(68, 75){\scriptsize \color{red}$d_2$}
		\put(75, 56){\scriptsize \color{red}$d_1$}
 		\put(19, 53){\scriptsize \rotatebox{0}{\color{blue}\shortstack{Line\\ Source}}}
 		\put(20, 46){\scriptsize \tjs{Tx}}
		\put(70, 38){\scriptsize \rotatebox{0}{\color{amber}\shortstack{Specular \\Transmission}}}
		\put(57, 67.5){\tiny \rotatebox{18}{ $m_1$}}
		\put(63, 38){\tiny \rotatebox{-40}{ $m_1$}}
		\put(63, 45){\tiny \rotatebox{50}{ $m_2$}}
		\put(60, 27){\tiny \rotatebox{-30}{ $m_0$}}
		\put(60, 61){\tiny \rotatebox{28}{ $m_0$}}
	\end{overpic} \caption{Simulation setup for the capturing the 2D fields from the metasurface $S$ of length $\ell_x=1\text{m}$ using Forward Ray Tracing. The source Tx is placed at a distance \ok{$\ell_z=0.5\text{m}$} away from the center of the metasurface. The resultant scattered rays are detected by equally spaced detectors at a radius of $R=1\text{m}$ from the origin, each with an angular length of $\Delta\phi$. Also shown are the specular scattered rays in \red{green,} red \red{and purple}, and the edge diffracted rays in \ok{blue}.\label{Fig:FRTSetup}
% 	\red{(YdJ: I suggest that we swap Figs. 2 and 4. This figure is useful for the reader to understand the forward ray tracing method introduced in Section III. The flowchart includes material that is introduced in this section, and therefore seems to be more at home here.)}
	}
\end{figure}
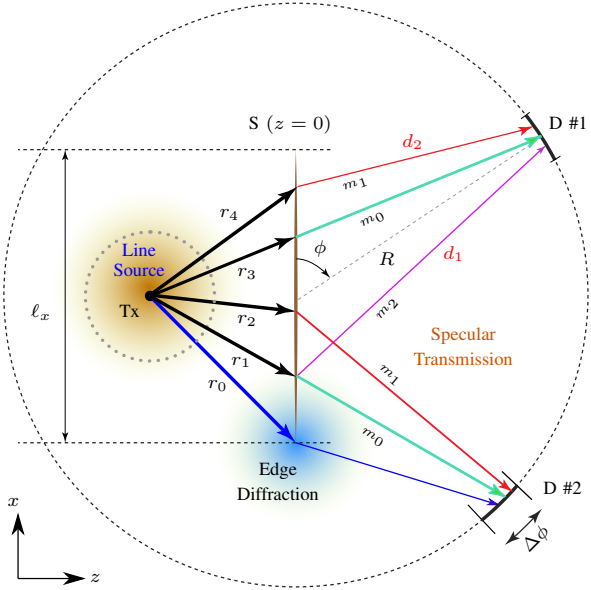

For $k\rightarrow\infty$, the surface integrals above reduce to discrete sums of scattering contributions from a set of critical points on $S$, resulting in a ray description of the scattered field \ok{(see Fig. \ref{Fig:FRTSetup})} \cite{pathak_UTD} .
Due to the 2D nature of the scattering geometry of Fig.~\ref{Fig:TxRx2}, \sg{only two kinds of critical points are considered here}.
Critical points of the first kind lie in the interior of $S$ and are associated with specularly reflected and transmitted fields. Their locations satisfy (\ref{eq:proj2}), a generalized version of Snell's laws of reflection and transmission, and are therefore dependent on $m$.
Critical points of the second kind, on the other hand, lie on the edges of $S$ and give rise to edge-diffracted fields.
Expressions for the scattered field amplitudes associated with each kind of critical point are provided in the following.

\subsubsection{Specular Fields}
An incident ray intersecting with the metasurface at an interior point $(x_c,0)$, with \mbox{$|x_c| \leq \ell_x/2$}, gives rise to specular scattering components, $\E^{(m)}_{s,a}$, with $a\in\{i,r,t\}$, whose scattering directions, $\hat{\bf k}_s$, satisfy
\begin{equation}
\hat{\bf k}_s =
\begin{cases}
    \hat{\bf k}_i       & \quad\text{if $a=i$ and $m=0$}\\
    \hat{\bf k}_a^{(m)} & \quad\text{if $a\in\{r,t\}$}.
    \nonumber
\end{cases}
\end{equation}
%
%\tjs{Critical points of the \nth{1} kind lie in the interior of $S$, for a ray scattering in the direction of $\khat_s$, if the following relations are satisfied:}
%
%\begin{align}\label{Eq:baRelation}
%    \begin{array}{ccc}
%        \b_a^{(m)}\cdot\uhat_1=0 & \text{and} & \b_a^{(m)}\cdot\uhat_2=0,
%    \end{array}
%\end{align}
%
%\noindent where $m$ can be any integer mode number. This expresses the tangential components of the wave-vectors of the incident and scattered waves following the phase matching condition at the metasurface location. For the incident fields with $a=i$ at $m=0$, $\khat_s=\khat_i = \khat_r^{(0)}$. In this case, it can be shown that the critical point contribution is equal in magnitude but the negative of the incident field at position $\p$, given by~\cite{yvo}
%, the latter of which follows Snell's law and whose contribution is equal to zero\cite{yvo}
If $a=i$, it can be shown \cite{yvo} that the critical point contribution is equal in magnitude but opposite to the incident field at the detection point, that is
\begin{align}\label{Eq:Shadow}
    \E_{s,i}^{(0)}(x,z) \simeq -\E_i(x_c) \sqrt{\frac{\rho_i}{\rho_i+s}} e^{-jks},
\end{align}
\noindent where $\rho_i$ is the radius of curvature of the incident wavefront, as in (\ref{Eq:Phii}), and $s$ is the distance from the critical point to a detector location at $(x,z)=(x_c,0) + s\hat{\bf k}_s$.
This field cancels the incident field in the transmission region (Region II) of the metasurface to create a shadow zone.
%\red{\textbf{What about its contribution in the reflection region, if $\p$ lies there?}} 
No shadow field exists in the reflection region (Region I). 
\ok{The field amplitudes corresponding to the specularly reflected and transmitted ray spectra are evaluated as}
\begin{subequations}\label{Eq:SpecularF}
    \begin{align}
        \E_{s,r}^{(m)}(x,z) &\simeq \E_r^{(m)}(x_c)
            \sqrt{\frac{\rho_r}{\rho_r+s}} e^{-jk\tilde{s}}\\
        \E_{s,t}^{(m)}(x,z) &\simeq \E_t^{(m)}(x_c)
            \sqrt{\frac{\rho_t}{\rho_t+s}} e^{-jk\tilde{s}}
    \end{align}
\end{subequations}
\ok{for $z \lessgtr 0$, respectively.}
Here, $\tilde{s}=s-m\psi(x_c)$ is the optical path length \cite{yvo} (corrected for the phase change induced by the metasurface), and
%\red{$\rho_r=\rho_t$}
\sg{
\begin{equation}
    \rho_r = \rho_t =
    \left\{
    \frac{1}{\rho_i} - m\ddot{\psi}(x_c)
    \right\}^{-1}
\end{equation}
}
\noindent is the radius of curvature of the reflected and transmitted wavefronts. 

\subsubsection{Edge-diffracted Fields}
A ray incident on an edge point $(x_c,0)$, with $x_c=\pm \ell_x/2$, produces a Keller cone \cite{balanis_2012} of diffracted field contributions
\begin{subequations}\label{Eq:EdgeF}
\begin{align}
    \E_{s,i}^{(0)}(x,z) &\simeq 
    \bar{\bf G}(\hat{\bf k}_i,\hat{\bf k}_s) \cdot \E_i(x_c)
    \ok{D^{(0)}_{e,i}(x_c)} \frac{e^{-jks}}{\sqrt{s}} \\
%\end{equation}
%\begin{equation}
    \E_{s,r}^{(m)}(x,z) &\simeq 
    \bar{\bf G}(\hat{\bf k}_r^{(m)},\hat{\bf k}_s) \cdot \E_r^{(m)}(x_c)
    \ok{D^{(m)}_{e,r}(x_c)} \frac{e^{-jks}}{\sqrt{s}} \\
%\end{equation}
%\begin{equation}
    \E_{s,t}^{(m)}(x,z) &\simeq 
    -\bar{\bf G}(\hat{\bf k}_t^{(m)},\hat{\bf k}_s) \cdot \E_t^{(m)}(x_c)
    \ok{D^{(m)}_{e,t}(x_c)} \frac{e^{-jks}}{\sqrt{s}}
\end{align}
\end{subequations}
to all detector locations $(x,z)$ in both Regions I and II; as previously, $s\hat{\bf k}_s$ is the vector from the critical point to the detector location.
%\tjs{Critical points of the \nth{2} kind lie on the edges }of the surface defined by two adjacent corner points and are characterized by two unit vectors: an edge-parallel vector $\ehat$ parallel with the straight edge, and a transverse vector $\that$ that is perpendicular to both $\ehat$ and $\nhat$ and points to the interior of $S$. \tjs{A critical point of the \nth{2} kind exists on such an edge if the following relationship is satisfied}
%
%\begin{align*}
%    \b_a^{(m)}\cdot\ehat = 0.
%\end{align*}
%
%The solutions for the scattering direction $\khat_s$ can be used to determine the diffraction point(s) given an observation point, or to calculate the half-angle of the Keller cone of scattered rays, given by \cite{balanis_2012} \cite{yvo}
%
%\begin{align*}
%    \beta=\sin^{-1}\left\{\sqrt{1-\left(\khat_s\cdot\ehat\right)}\right\}
%\end{align*}
%
%The scattered diffracted field of each mode $m$, can be evaluated as:
%
% \begin{align*}\label{Eq:EdgeF}
%     \E_{s,a}^{(m)}(\p)\approx\bar{\mathbf{D}}_{e,a}^{(m)}\cdot\E_i(\q_c)\sqrt{\frac{\rho_{e,a}}{s\left(\rho_{e,a}+s\right)}}e^{-jks}
% \end{align*}
%\tjs{
%\begin{subequations}\label{Eq:EdgeF}
%\begin{align}
%    \E_{s,i}^{(0)}(\p)&\approx D_{e,i}(s) \bar{\G}(\hat{\mathbf{k}}_i,\q_c)\cdot\E_i(\q_c) R_i(s) \\
%    \E_{s,r}^{(m)}(\p)&\approx D_{e,r}(s) \bar{\G}(\hat{\mathbf{k}}_r^{(m)},\q_c)\cdot\E_r(\q_c) R_r(s) \\
%    \E_{s,t}^{(m)}(\p)&\approx -D_{e,t}(s) \bar{\G}(\hat{\mathbf{k}}_t^{(m)},\q_c)\cdot\E_t(\q_c) R_t(s) 
%\end{align}
%\end{subequations}
Furthermore,
\begin{equation}
    \ok{D^{(m)}_{e,a}} = \frac{F(X_e)}{2\sqrt{2\pi jk}
    \ok{({\bf b}^{(m)} \cdot \hat{\bf t})}}
    %  \quad \text{and} \quad
%    R_a(s)  &= \sqrt{\frac{\rho_{e,a}}{s\left(\rho_{e,a}+s\right)}}e^{-jks}
\end{equation}
is an edge-diffraction coefficient in which
\ok{
\begin{equation}
    {\bf b}^{(m)} = \hat{\bf k}_i - \hat{\bf k}_s
    - m\dot{\psi}(x_c) \, \hat{\bf x}
\end{equation}
and $\hat{\bf t}$ is perpendicular to the edge and points into the interior of $S$, i.e., $\hat{\bf t}=\mp\hat{\bf x}$ at $x_c=\pm\ell_x/2$.
Finally, $F(X_e)$ is the UTD transition function with argument
\begin{equation}
    X_e = \frac{ks}{2}
    \left( \frac{{\bf b}^{(m)} \cdot \hat{\bf t}}{\hat{\bf k}_s \cdot \hat{\bf n}} \right)^2,
\end{equation}
}
which ensures the total scattered field to be continuous across shadow boundaries \cite{yvo}.

\section{SSFT based Determination of $\psi$ and $\bar{\bm\chi}^{(m)}$}\label{Sec:ParametricVariable}
As shown in Sec.~\ref{Eq:GSTC_Solution}, the surface fields (\ref{Eq:ScatteredFieldEquations}) needed to compute the scattered fields away from the metasurface can be determined under the assumption that the surface susceptibilities are periodic functions (\ref{Eq:ChiFourierSeries}) of the parameter $\psi$.
In (\ref{Eq:ChiFourierSeries}), $\bar{\bm\chi}^{(m)}_{ab}(x')$ denotes the amplitude of the $m^{th}$ mode of $\bar{\bm\chi}_{ab}$ near $x'$, which has a localized spatial frequency $km\dot{\psi}(x')$.
%in the Fourier expansion form,
%
%\begin{align}\label{Eq:iFourierSeries}
%    \bar{\chi}_{ab}(\psi;\q') = \sum_{m=-\infty}^{\infty}\bar{\chi}_{ab}^{(m)}(\q')e^{jkm\psi},
%\end{align}
%%
%in terms of the parametric variable $\psi(\q')$, where $km\psi$ represents the localized spatial frequency contents of $ \bar{\chi}_{ab}$ at $\q=\q'$ in the $\psi-$space, and $k = 2\pi/\lambda$ is the wavenumber. This would capture the local periodicity of the surface, and make it integrable into the GSTCs for subsequently computing the scattered fields.
Our objective now is to estimate these location-dependent spatial frequencies across an arbitrary given surface susceptibility profile, and
determine the corresponding modal amplitudes.
%express them in the form \eqref{Eq:iFourierSeries}.
To this end, we investigate the \emph{Short Time Fourier Transform (STFT)}, conventionally applied to time-varying non-stationary signals \cite{KEHTARNAVAZ2008175}, as a means to estimate the localized spatial frequency content of $\bar{\bm\chi}_{ab}$.
When applied to the problem considered herein, we analogously refer to this procedure as the \emph{Short Space Fourier Transform (SSFT)}.

To describe the procedure, assume for the moment that the metasurface is characterized by a scalar susceptibility that is locally periodic, and can therefore be expressed as
\begin{equation}\label{Eq:ChiScalar}
    \chi(x) = \sum_m \chi^{(m)}(x') e^{jkm\psi(x)},
\end{equation}
where $\psi(x)$ is well approximated by (\ref{Eq:PsiQuadratic})
if $x$ is sufficiently close to $x'$.
The SSFT of $\chi$, denoted by $\mathcal{S}\{\chi(x)\}=\mathcal{S}(\kappa, x')$, involves computing the Fourier transform of $\chi$ multiplied by a space-limited window function $w(x-x')$ centered at $x=x'$, which slides along the $x$-axis;
$\kappa$, in rad/m, denotes the spatial frequency.
At each window location,
\begin{equation}\label{Eq:SSFT}
    \mathcal{S}(\kappa, x')
    = \int_{-\infty}^\infty
    \chi(x)w(x-x')e^{-j\kappa x}dx
\end{equation}
represents the localized spectral content of the profile.
Substituting (\ref{Eq:ChiScalar}) and (\ref{Eq:PsiQuadratic}), it is not difficult to show that $\mathcal{S}\{\chi\}$ is well approximated by
\begin{equation}\label{Eq:SSFT2}
    \mathcal{S}(\kappa, x') \simeq e^{-j\kappa x'}
    \sum_m \chi^{(m)}(x') e^{jkm\psi(x')}
    W(\kappa - km\dot{\psi}),
\end{equation}
%\begin{equation}\label{Eq:SSFT2}
%    \mathcal{S}(\kappa, x') \sim
%    \sum_m \chi^{(m)}(x') e^{jkm\psi(x')}
%    W(\kappa - km\dot{\psi}),
%\end{equation}
provided that the window function is narrow enough for the quadratic term in (\ref{Eq:PsiQuadratic}) to be negligibly small for all $x$ where $w(x-x') \neq 0$.
In (\ref{Eq:SSFT2}),
\begin{equation}
    W(\kappa) = \int_{-\infty}^{\infty}
    w(x) e^{-j\kappa x} dx \notag
\end{equation}
is the window function in the spatial Fourier domain.

\begin{figure}[!t]
\centering
	\begin{overpic}[grid=false, width=0.8\columnwidth]{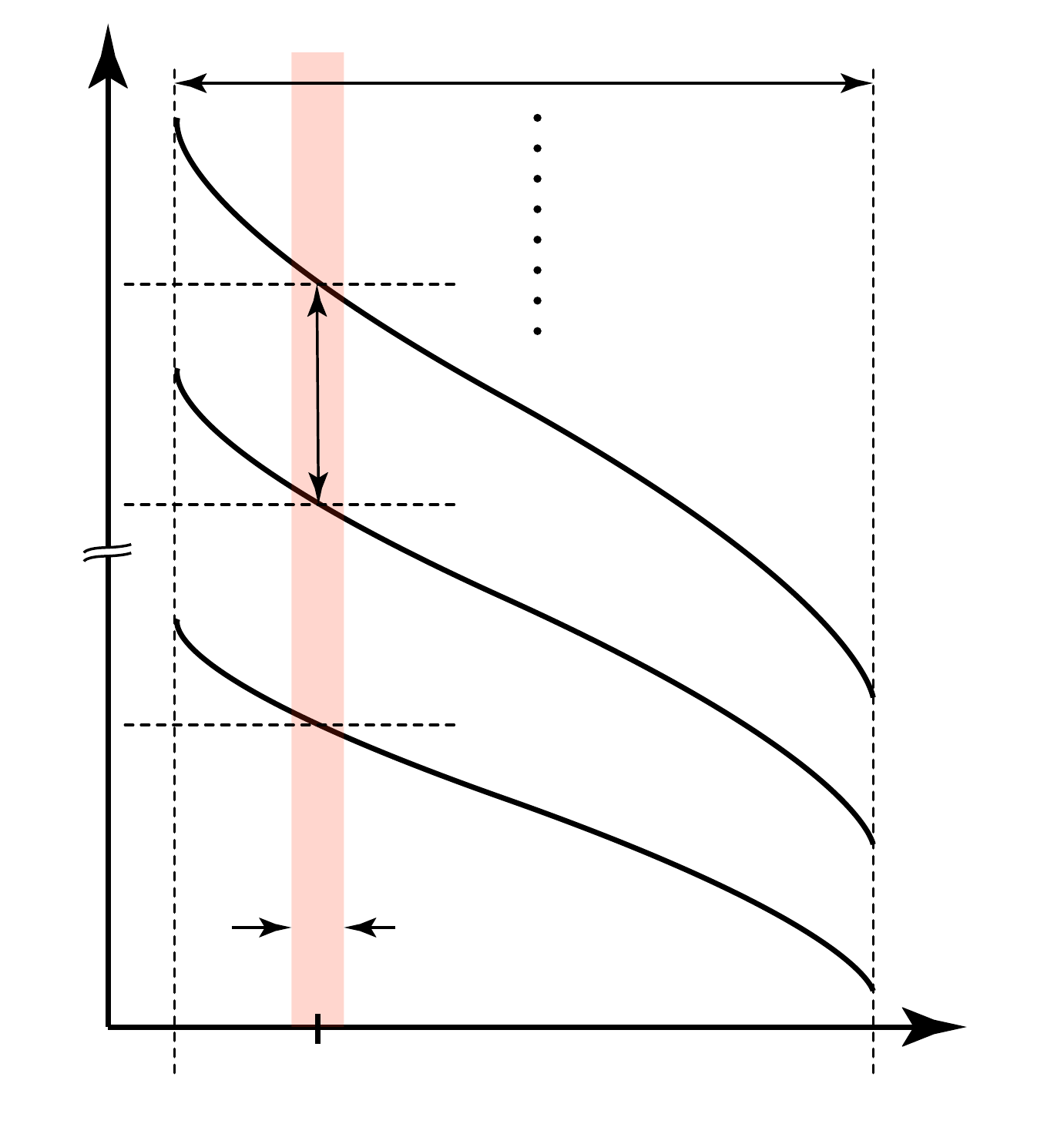}
			\put(50,5){ \makebox(0,0){\scriptsize Surface Location, $x$}}
			\put(4, 50){ \rotatebox{90}{\scriptsize \makebox(0,0){Spatial Frequency, $\kappa$}}}
			\put(27,5){ \makebox(0,0){\scriptsize $x'$}}
			\put(49,18){ \makebox(0,0){\scriptsize Window, $w(x-x')$}}
			\put(55, 30){ \scriptsize\makebox(0,0){$m-1$}}
			\put(55, 45){ \scriptsize\makebox(0,0){$m$}}
			\put(55, 64){ \scriptsize\makebox(0,0){$m+1$}}
			\put(50,95){\scriptsize \makebox(0,0){Metasurface Size, $\ell_x$}}
			\put(30,66){ \scriptsize\makebox(0,0){$\Delta\kappa$}}
			\put(44, 36){ \scriptsize\makebox(0,0){$\kappa_{m-1}$}}
			\put(42, 55){ \scriptsize\makebox(0,0){$\kappa_m$}}
	\end{overpic}
    \caption{An illustration of a typical Short Space Fourier Transform (SSFT) distribution exhibiting uniformly spaced discrete spatial frequencies following local periodicities of the surface, which are varying over the surface.}\label{Fig:Tech}
\end{figure}
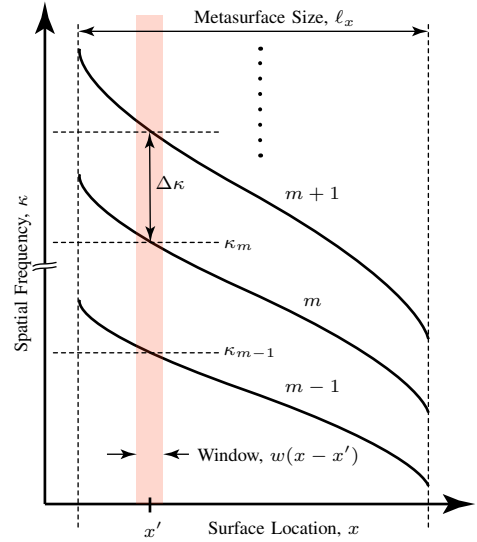
The SSFT distribution of a locally periodic susceptibility profile typically looks like that of
%For a locally periodic metasurface, the typical SSFT profile may look like that of
Fig.~\ref{Fig:Tech}, which shows discrete bands of high amplitude that are uniformly spaced along $\kappa$, with a spacing that varies with the window location $x'$.
Recognizing that these bands correspond with the summation terms of (\ref{Eq:SSFT2}), their spacing, $\Delta\kappa$, equals $k\dot{\psi}$ and the locations of their maxima with respect to $\kappa$, denoted by $\kappa_m$, can be used to estimate the local periodicity of the metasurface as follows:
\begin{align}\label{Eq:g_psi_SSFT}
    \dot{\psi} \simeq \frac{\kappa_m}{km}.
\end{align}
Having estimated $\dot{\psi}$ for a sufficiently dense grid of surface points, the phase function $\psi(x)$ can be numerically constructed via
\begin{align}\label{Eq:psi_SSFT}
    \psi(x)= \psi(-\ell_x/2) + \int_{-\ell_x/2}^{x} \dot{\psi}(x') dx',
\end{align}
where the constant of integration can be chosen arbitrarily, and is set to zero for simplicity here.
% Finally, the unknown ${\chi}^{(m)}$ around $x'$ can be estimated from the SSFT distribution at $\kappa=\kappa_m$, i.e.,
% %direct numerical integration of ${\chi}$ using the estimated $\psi(x)$ from \eqref{Eq:psi_SSFT}, i.e.
% %
% \begin{align}\label{Eq:ExpCoeff}
%     {\chi}^{(m)}(x') \simeq
%     \frac{\mathcal{S}(\kappa_m, x')}{W(0)}
%     e^{j\kappa_m x'} e^{-jkm\psi}.
% \end{align}
%
%\begin{align}\label{Eq:ExpCoeff}
%    \bar{\chi}_{ab}^{(m)}(x) = %\frac{1}{\lambda}\int_0^\lambda %\bar{\chi}_{ab}(\psi;x)e^{-jkm\psi}d\psi.
%\end{align}

\ok{Generalizing this procedure to a set of dyadic surface susceptibility profiles $\bar{\bm\chi}_{ab}(x)$, with $a\in\{{\rm e}, {\rm m}\}$, the gradient (slope) function $\dot{\psi}(x)$ is first determined through \eqref{Eq:g_psi_SSFT} using the localized spatial frequencies estimated from a dyadic version of the SSFT integral \eqref{Eq:SSFT}.
The phase function $\psi(x)$ is next computed using \eqref{Eq:psi_SSFT} using with a polynomial curve fitted to the extracted $\dot \psi$ data.}
% , followed by estimates of the Fourier coefficients $\bar{\bm\chi}_{ab}^{(m)}$ using a dyadic version of \eqref{Eq:ExpCoeff}.
Care must be taken that the assumption of linearly varying phase within the spatial limits of the window function $w(\cdot)$ is satisfied.
\ok{Finally, each unknown \red{$\bar{\bm\chi}_{ab}^{(m)}(x)$} may be obtained by direct numerical integration of \red{$\bar{\bm\chi}_{ab}(x)$} (see \eqref{eq:FourierCoeff}), using \red{$\bar{\bm\chi}_{ab}(\psi;x') \simeq \bar{\bm\chi}_{ab}(x)$} and the estimated $\psi(x)$ from \eqref{Eq:psi_SSFT},}
%
% \begin{align}\label{Eq:ExpCoeff}
%     \bar{\chi}_{ab}^{(m)}(x) = \frac{1}{\lambda}\int_0^\lambda \bar{\chi}_{ab}(\psi;x)e^{-jkm\psi}d\psi.
% \end{align}}
\ok{
\begin{align*}
    \bar{\bm\chi}_{ab}^{(m)}(x') &= \frac{1}{\lambda}\int_{\psi_1}^{\psi_2} \bar{\bm\chi}_{ab}(x)e^{-jkm\psi}d\psi,
\end{align*}
where $\psi_{1,2}$ \red{lie} on either side of $\psi(x')$ and span one full period such that $\psi_2 - \psi_1=\lambda$. We can then use the relationship $d\psi = \dot{\psi}dx$ to change the integration to be over $x$ and we obtain,
\begin{align} \label{eq:chiCoeff}
    \bar{\bm\chi}_{ab}^{(m)}(x') &= \frac{1}{\lambda}\int_{x_1}^{x_2} \bar{\bm\chi}_{ab}(x)e^{-jkm\psi(x)}\dot{\psi}(x)dx
\end{align}
with the limits given by $x_{1,2}=\psi^{-1}(\psi_{1,2}$).
}

It is conceptually straightforward to extend the procedure to the wider class of 2D (locally periodic) metasurfaces, whose susceptibility profiles vary along the direction of a gradient vector that is not necessarily parallel to $\hat{\bf x}$.
To do this, a 2D spatial Fourier transform needs to be applied to $\bar{\bm\chi}_{ab}(x, y)$ multiplied by a 2D window function.
The locations $(\kappa_{1,m},\kappa_{2,m})$ of the spectral maxima at a given surface location correspond to a 2D spatial frequency vector ${\bm\kappa}_m=\kappa_{1,m}\uhat_1 + \kappa_{2,m}\uhat_2$, from which one can construct the phase function $\psi(x,y)$ and its gradient ${\bf g}_{\psi}(x,y)$.

While the above procedure is simple, it also has limitations.
The SSFT integral \eqref{Eq:SSFT} suffers from the well-known uncertainty principle, according to which its resolution (the minimum distance at which nearby space-frequency components can be resolved) is inversely related along $\kappa$ and $x$.
Resolution in one domain can be traded off for increased resolution in the other by choosing a different window function $w(\cdot)$, but in some cases it may not be possible to find a window function that avoids any significant space-frequency overlap. \ok{However, as we shall see, the use of a fitted function to represent $\dot \psi$ alleviates these issues if discretion is taken in the retention of points in areas where the SSFT cannot discriminate the modal frequencies clearly.}

% In such cases it may be possible to utilize more advanced, so-called superresolution techniques \cite{?} \red{(add reference or remove this comment)} that are outside the scope this paper.

To demonstrate the utility of the procedure we consider the case of a horizontally oriented metasurface lens of size \mbox{$\ell_x=1$~m} (see Fig. \ref{Fig:FRTSetup}) designed to collimate a line source at $(0,-f)$, with $f=0.5$~m, to a normally outgoing plane wave propagating along $z > 0$.
This requires the metasurface to have a hyperbolic phase profile~\cite{voelz_2010}
%To synthesize a susceptibility profile for such a surface, consider the hyperbolic phase profile of an optical lens in the $x$--$z$ plane that images from point ($x_0$, $y_0$, $z_0$) to point ($x_i$, $y_i$, $z_i$)~\cite{voelz_2010}:
%
\begin{align*}
    \phi(x) =& -\frac{2\pi f}{\lambda} \left(\sqrt{\frac{x ^2}{f^2}+1} - 1\right). %\label{Eq:LensPhaseProfile2}
\end{align*}
% \blue{and a loss profile given by,
% \begin{align*}
%     ???
% \end{align*}}
%
% Now assume that the incident field has a frequency of 60 GHz, TM$^y$ polarization, and unit magnitude.
Now assume that the incident field \ok{is a cylindrical wave with a frequency of 60 GHz, with TM$^y$ polarization originating from $(0,-f)$. If normalized so that the clylindrical wave has a value of ${\bf E}_i(0,0) = 1\hat{\bf y}$ at the center of the surface, the field can be given by}
\begin{align*}
    \E_i(x,z) &= \frac{M_f\omega\mu_0}{4}H_{0}^{(2)}\left(k\sqrt{x^2 + (z+f)^2}\right)\hat{\bf y}\\
    \H_i(x,z) &= \frac{-ikM_f\left(z+f\right)}{4\sqrt{x^2 + (z+f)^2}}H_{1}^{(2)}\left(k\sqrt{x^2 + (z+f)^2}\right)\hat{\bf x}
\end{align*}
\noindent\ok{where $H_\nu^{(2)}(X)$ is the Hankel function of the second kind and $M_f = 4/(\omega\mu_0 H_0^{(2)}(k|f|))$ is a normalization factor.} 

% \red{Regarding the assumption that the incident field has unit magnitude: As it is cylindrical, its magnitude on the surface is dependent on $x$. At the source it cannot be equal to one, either. I will leave the solution to this problem to Scott.}
% \blue{The metasurface is assumed to be perfectly matched (${\bf E}_r = {\bf 0}$) and transmit a surface field ${\bf E}_t(x)= 0.4j e^{jk\phi(x)}\hat{\bf y}$ (\red{same issue}) on $S_{-}$.
% \red{We need to state our synthesis constraints here: we assume that the surface is isotropic and can be described by tangential components of $\bar{\bm\chi}_{\rm ee}$ and $\bar{\bm\chi}_{\rm mm}$ only.}
% Using the local plane-wave approximation, the corresponding magnetic surface field on $S_{-}$ is given by }
% % \red{\textbf{[It would have been ideal to rigorously synthesize the surfaces for all the examples as approximated fields in the synthesis stage is unnecessary.]}}
% %
% \begin{align*}
% H_r(x,\tjs{y}) \simeq \cos(\theta_t)\frac{e^{j\phi(x)}e^{j\beta_{\tjs{y}}\tjs{y}}}{\eta}~\hat{x}
% \end{align*}
% %

\begin{figure}[!t]
\centering
	\begin{overpic}[grid=false, width=\columnwidth]{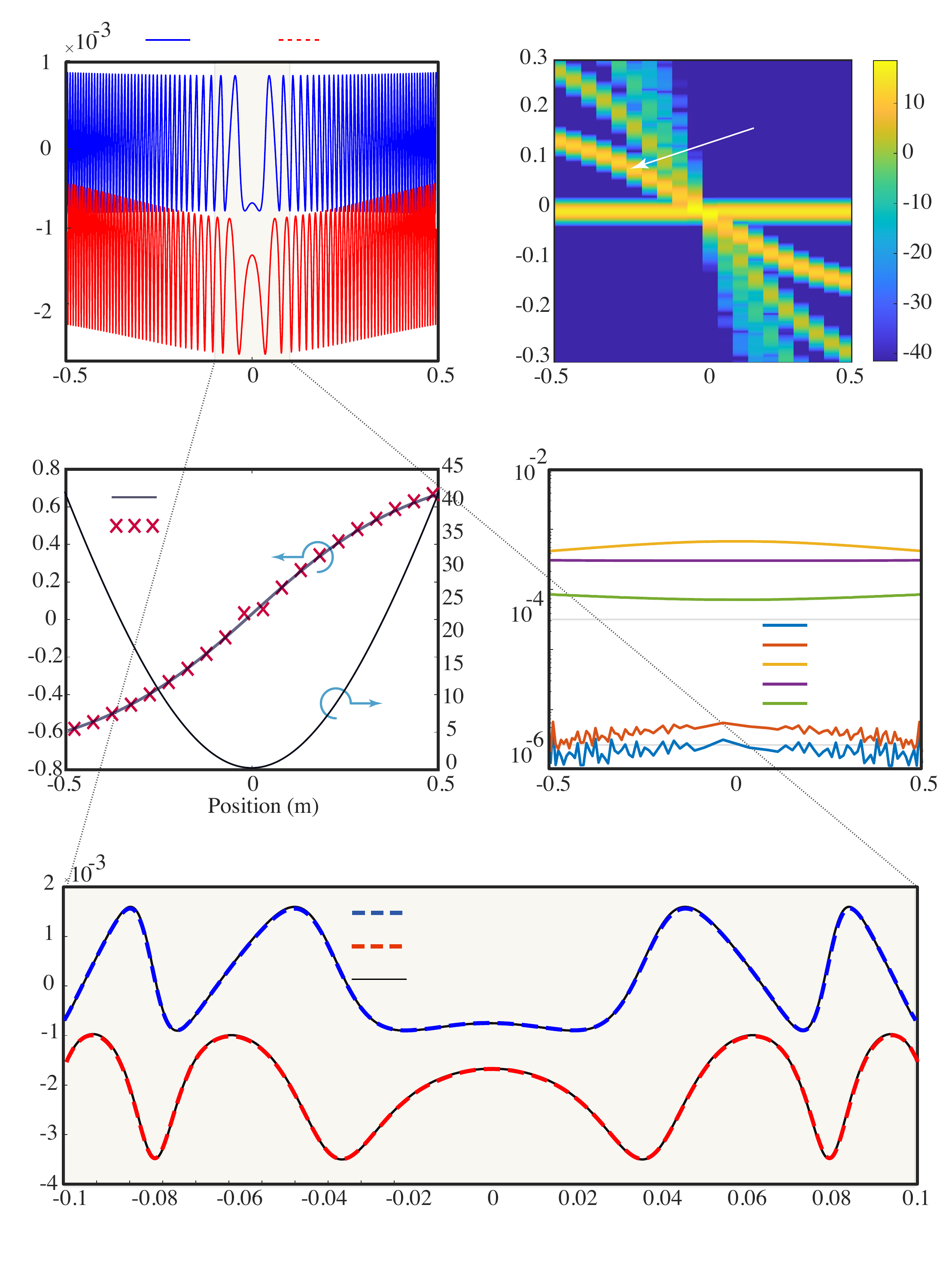}
			\put(17, 68){ \makebox(0,0){\scriptsize Position, $x$~(m)}}
			\put(37, 3){ \makebox(0,0){\scriptsize Position, $x$~(m)}}
			\put(37, 0){ \makebox(0,0){\scriptsize (e)}}
			\put(19, 65.5){ \makebox(0,0){\scriptsize (a)}}
			\put(19, 33){ \makebox(0,0){\scriptsize (c)}}
			\put(0, 84){ \rotatebox{90}{\scriptsize \makebox(0,0){susceptibility, $\chi_\text{ee}(x)$}}}
			\put(0, 18){ \rotatebox{90}{\scriptsize \makebox(0,0){susceptibility, $\chi_\text{ee}(x)$}}}
			\put(0, 51){ \rotatebox{90}{\scriptsize \makebox(0,0){gradient function, $\ok{\dot \psi}$}}}
%			\put(37, 50.5){ \rotatebox{-90}{\scriptsize \makebox(0,0){parametric variable, $\psi$, \eqref{Eq:psi_SSFT}}}}
			\put(37, 50.5){ \rotatebox{-90}{\scriptsize             \makebox(0,0){phase function, $\psi$, \eqref{Eq:psi_SSFT}}}}
			\put(74, 50.5){ \rotatebox{-90}{\scriptsize \makebox(0,0){Coefficients, $\chi_\text{ee}^{(m)}$, \eqref{eq:chiCoeff}}}}%\eqref{Eq:ExpCoeff}}}}
			\put(17, 97){ \makebox(0,0){\tiny Re$\{\cdot\}$}}
			\put(27, 97){ \makebox(0,0){\tiny Im$\{\cdot\}$}}
			\put(17, 60.9){ \makebox(0,0){\tiny fitted $\ok{\dot \psi}(x)$}}
			\put(17, 58.8){ \makebox(0,0){\tiny $\ok{\dot \psi}(x)$, \eqref{Eq:g_psi_SSFT}}}
			\put(55, 68){ \makebox(0,0){\scriptsize Window Position, $x'$~(m)}}
			\put(38, 83){ \rotatebox{90}{\scriptsize \makebox(0,0){ frequency, $\kappa$~(rad/m)}}}
			\put(55, 98){ \makebox(0,0){\scriptsize $|\mathcal{S}\{\chi_\text{ee}(x)\}| = |\mathcal{S}(\kappa,x')|$~(dB)}}
			\put(62, 90.4){ \makebox(0,0){\color{white}\scriptsize $m=1$}}
			\put(55, 65.5){ \makebox(0,0){\scriptsize (b)}}
			\put(57, 33){ \makebox(0,0){\scriptsize (d)}}
			\put(57, 36){ \makebox(0,0){\scriptsize Position, $x$~(m)}}
%			\put(84, 31){ \makebox(0,0){\scriptsize (c)}}
			\put(66, 50.5){ \makebox(0,0){\tiny m = -2}}
			\put(66, 49){ \makebox(0,0){\tiny m = -1}}
			\put(66, 47.5){ \makebox(0,0){\tiny m = 0}}
			\put(66, 46){ \makebox(0,0){\tiny m = +1}}
			\put(66, 44.5){ \makebox(0,0){\tiny m = +2}}
			\put(35, 28){ \makebox(0,0){\scriptsize $\Re[\chi_\text{ee}]$} }
			\put(35, 25.2){ \makebox(0,0){\scriptsize $\Im[\chi_\text{ee}]$} }
			\put(37, 22.2){ \makebox(0,0){\scriptsize original $\chi_\text{ee}$}}
			\put(39, 26.5){ \makebox(0,0){{\Large \}}}}
			\put(45, 26.4){ \makebox(0,0){\scriptsize From \eqref{Eq:ChiScalar}}}
	\end{overpic}
    \caption{The procedure to convert a specified surface susceptibility profile of a metasurface in a Fourier \ok{series form of \eqref{Eq:ChiScalar}}, illustrated using an example of a reflective metasurface. (a) The electric surface susceptibility profile, $\chi_\text{ee}(x)$ across the surface. b) The SSFT magnitude using a Kaiser window, \eqref{Eq:SSFT}. c) The extracted and fitted gradient function $\ok{\dot \psi}$, and the corresponding parametric variable $\psi$ using \eqref{Eq:g_psi_SSFT}. d) Corresponding expansions coefficients, $\chi_\text{ee}^{(m)}$ for few indices $m$. e) The reconstructed surface susceptibility using the expansion \ok{form of \eqref{Eq:ChiScalar}}, compared to the original in a smaller spatial region for clarity.}\label{Fig:SSFT_Exp}
\end{figure}
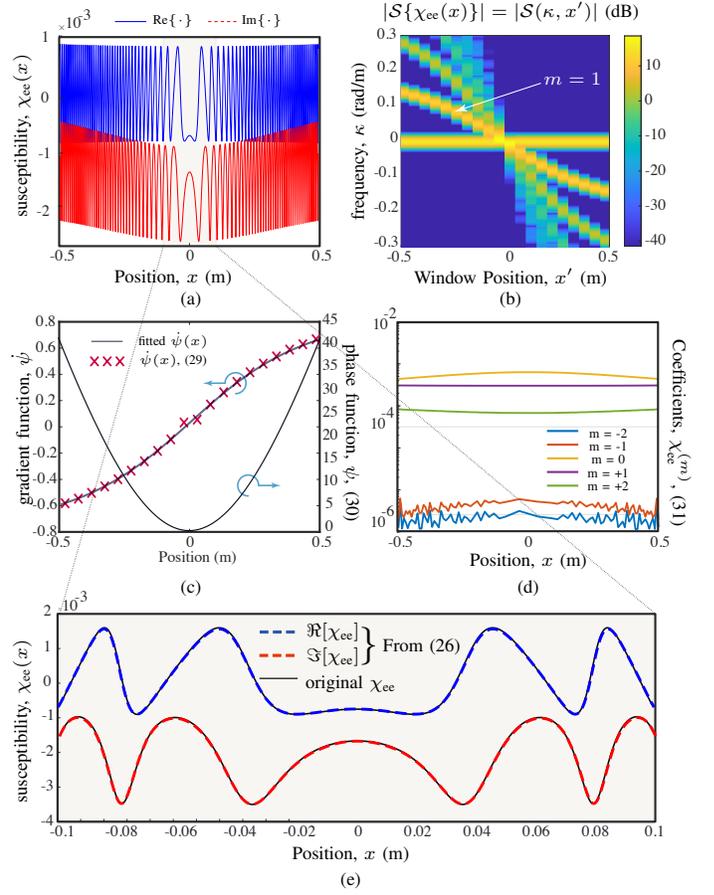

\ok{The metasurface is assumed to be perfectly matched (${\bf E}_r = {\bf 0}$) and transmit an electric and magnetic surface field ${\bf E}_t(x)= 0.2j \hat{\bf y}$ and ${\bf H}_t(x)= -0.2j/\eta \hat{\bf x}$, where $\eta$ is the free space impedance. The metasurface therefore introduces a position dependent loss that accommodates the variation in $|\E_i|$ along the surface to produce a uniform transmitted field intensity. If we constrain the surface synthesis to an isotropic surface, described by the tangential components of $\bar{\bm\chi}_{\rm mm}$ only, the corresponding electric and magnetic surface susceptibilities are then obtained using:}

% where $\eta$ is the free space impedance.
%The incident fields are given by $E_0 = 1\:\tjs{\hat{z}}$ and $H_0 = 1/\eta~\hat{x}$.
% The corresponding electric and magnetic surface susceptibilities are then obtained using:
%
\begin{align*}
\chi_\text{ee}(x) &= \frac{\Delta H_{\ok{x}}}{j\omega \epsilon_0E_{z,\text{avg}}}\\
\chi_\text{mm}(x) &= \frac{\Delta E_{\ok{y}}}{j\omega \epsilon_0H_{\tjs{x},\text{avg}}}
\end{align*}
where $\Delta\{\cdot\}$ is the field difference across the metasurface, and $\{\cdot\}_\text{avg}$ is the average fields \cite{metasurface_synthesis}. For this example, the computed $\chi_\text{ee}(x)$ ($\chi_\text{mm}$ not shown for conciseness) is shown in Fig.~\ref{Fig:SSFT_Exp}(a).

\tjs{In order to express the susceptibilities in the form of (\ref{Eq:ChiFourierSeries}), we first compute the SSFT using MATLAB's STFT function,
% \footnote{Here we are, of course, implicitly using a discrete form of \eqref{Eq:iFourierSeries} based on an FFT.}
the results of which are shown in Fig.~\ref{Fig:SSFT_Exp}(b).}
The window function $w(p)$ used was the discrete Kaiser window~\cite{kaiser}\cite{smith_2011}.
\ok{The surface was discretized with a spacing of $\lambda/400$ and the spatial extent of the window is approximately 0.20~m. 
% The total size of the FFT was $N \times 8$.
As can be seen, the SSFT decomposes the surface susceptibility into distinct spatial modes for multiple positions along the surface. Using the $m=1$ mode, the spatial frequencies are then extracted, and used in \eqref{Eq:g_psi_SSFT} to compute \ok{$\dot{\psi}$}, which is shown in Fig.~\ref{Fig:SSFT_Exp}(c). Notice that for this particular example, it is difficult to extract $\kappa_m$ near $x=0$ due to limited spatial and frequency resolution.}
% To obtain a smooth function over space, \tjs{a polynomial curve is fitted to the extracted data} which is also shown in Fig.~\ref{Fig:SSFT_Exp}(c).
% \red{(I'm inclined to move this part about polynomial curve fitting to an earlier paragraph where we discuss the limitations of the SSFT, but I'll wait to hear what the others think.)}

Next, \ok{$\dot{\psi}$} is integrated \ok{using a polynomial fit} following \eqref{Eq:psi_SSFT} and $\psi$ is obtained as shown in Fig.~\ref{Fig:SSFT_Exp}(c).
\ok{The last step is to compute \ok{the Fourier coefficients for $\chi_{\rm ee}$ and $\chi_{\rm mm}$} \red{using \eqref{eq:chiCoeff}}; those for $\chi_{\rm ee}$ are shown for a few modes in Fig.~\ref{Fig:SSFT_Exp}(d).
Fig.~\ref{Fig:SSFT_Exp}(e) compares the original profile of $\chi_{\rm ee}$ with a reconstructed profile based on its estimated Fourier coefficients.}
%\eqref{Eq:iFourierSeries} form is plotted using the extracted $\psi$ and $\chi_\text{ee}^{(m)}$, which is shown and compared with the original susceptibility function in Fig.~\ref{Fig:SSFT_Exp}(e).
An excellent match confirms the accuracy of the proposed method.
%that the susceptibility conversion to Fourier expansion form has been successfully completed.
%A similar process can be applied to the other surface susceptibility profiles.

\section{\red{Ray-Optical \sg{(RO)}-GSTC} Demonstration}

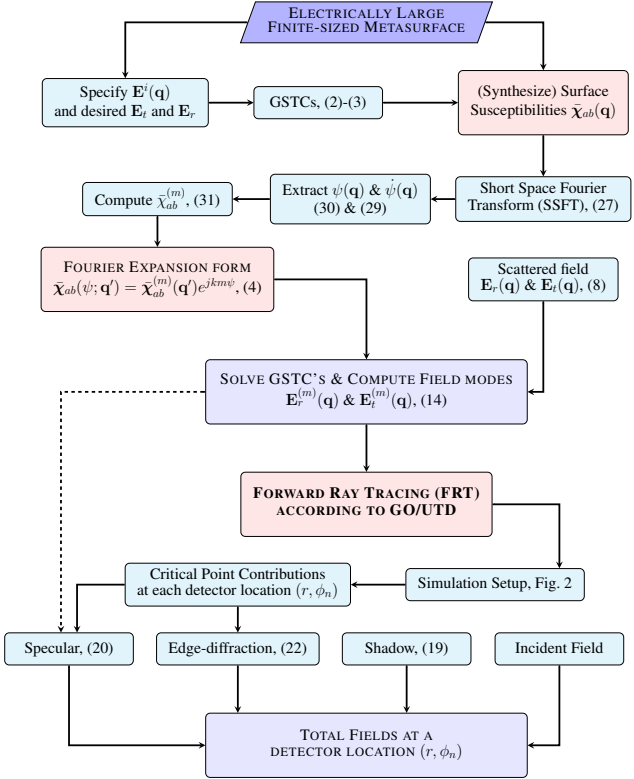
\begin{figure}[htbp]
%\begin{center}
\resizebox{\columnwidth}{!}{ 
\tikzstyle{startstop} = [rectangle, rounded corners, minimum width=3cm, minimum height=1cm,text centered, draw=black, fill=cyan!50]
\tikzstyle{io} = [trapezium, trapezium left angle=70, trapezium right angle=110, minimum width=3cm, minimum height=1cm, text centered, draw=black, fill=blue!30]
\tikzstyle{process} = [rectangle, rounded corners, minimum width=4cm, minimum height=1cm, text centered, draw=black, fill=cyan!10]
\tikzstyle{Main} = [rectangle, rounded corners, minimum width=4cm, minimum height=2cm, text centered, draw=black, fill=red!10]
\tikzstyle{Solve} = [rectangle, rounded corners, minimum width=10cm, minimum height=2cm, text centered, draw=black, fill=blue!10]

\tikzstyle{decision} = [diamond, rounded corners, minimum width=3cm, minimum height=1cm, text centered, draw=black, fill=cyan!30]
\tikzstyle{arrow} = [ultra thick, black, round cap->,>=stealth]
\tikzstyle{arrow2} = [ultra thick, dashed, black, round cap->,>=stealth]
\tikzstyle{line} = [draw, -latex']
%\clearpage
\centering
\begin{tikzpicture}[node distance=2cm] 
\Large\node (start) [io] {\begin{tabular}{c}\textsc{\shortstack{Electrically Large \\ Finite-sized Metasurface}}\end{tabular}};\Large

\Large\node (A1) [process, below of= start, yshift=-0.5cm, xshift=-7.5cm] {\begin{tabular}{c}\shortstack{Specify $\E^i(\q)$ \\and desired $\E_t$ and $\E_r$}\end{tabular}};\Large
\Large\node (A2) [process, right of= A1, yshift=-0cm, xshift=4cm] {\begin{tabular}{c}\shortstack{GSTCs, \eqref{Eq:GSTCs}-\eqref{Eq:PMs}}\end{tabular}};\Large
\Large\node (A3) [Main, right of= A2, yshift=0cm, xshift= 5cm] {\begin{tabular}{c}\shortstack{(Synthesize) Surface \\Susceptibilities $\bar{\bm\chi}_{ab}(\q)$}\end{tabular}};\Large
\Large\node (A4) [process, below of= A3, yshift= -1cm, xshift= 0 cm] {\begin{tabular}{c}\shortstack{ Short Space Fourier \\Transform (SSFT), \eqref{Eq:SSFT}}\end{tabular}};\Large
\Large\node (A5) [process, left of= A4, yshift= 0cm, xshift= -4 cm] {\begin{tabular}{c}\shortstack{ Extract $\psi(\q)$ \& $\dot \psi(\q)$\\ \eqref{Eq:psi_SSFT} \& \eqref{Eq:g_psi_SSFT}}\end{tabular}};\Large
\Large\node (A6) [process, left of= A5, yshift=-0cm, xshift= -4 cm]{\begin{tabular}{c}\shortstack{Compute $\bar{\chi}_{ab}^{(m)}$, \eqref{eq:chiCoeff}}\end{tabular}};\Large
\Large\node (A7) [Main, below of= A6, yshift=-0.5cm, xshift= 0 cm]{\begin{tabular}{c}\shortstack{\textsc{Fourier Expansion form}\\ $\bar{\bm\chi}_{ab}(\psi;\q') = \bar{\bm\chi}_{ab}^{(m)}(\q')e^{jkm\psi}$,~\eqref{Eq:ChiFourierSeries}}\end{tabular}};\Large
\Large\node (A8) [process, right of= A7, yshift=0cm, xshift= 10cm]{\begin{tabular}{c}\shortstack{Scattered field \\$\E_r(\q)$~\&~$\E_t(\q)$,~\eqref{Eq:ScatteredFieldEquations}}\end{tabular}};\Large
\Large\node (A9) [Solve, below of= start, yshift=-9.5cm, xshift= 0 cm]{\begin{tabular}{c}\shortstack{\textsc{Solve GSTC's \& Compute Field modes} \\$\E_r^{(m)}(\q)$~\&~$\E_t^{(m)}(\q)$,~\eqref{Eq:LinearFieldEquations}}\end{tabular}};\Large
\Large\node (A10) [Main, below of= A9, yshift=-1.5cm, xshift= 0 cm, label={[font=\large\sffamily,name=label1] }]{\begin{tabular}{c}\textbf{\textsc{\shortstack{Forward Ray Tracing (FRT) \\according to GO/UTD}}}\end{tabular}};\Large
\Large\node (A11) [process, below of= A10, yshift=-0.5cm, xshift= 4 cm]{\begin{tabular}{c} \shortstack{Simulation Setup,~Fig.~\ref{Fig:FRTSetup}}\end{tabular}};\Large
\Large\node (A12) [process, below of= A10, yshift=-0.5cm, xshift= -4 cm]{\begin{tabular}{c} \shortstack{Critical Point Contributions\\ at each detector location $(r, \phi_n)$}\end{tabular}};\Large
\Large\node (A15) [process, below of= A11, yshift=-0cm, xshift= -13.25 cm]{\begin{tabular}{c} \shortstack{Specular, \eqref{Eq:SpecularF}}\end{tabular}};\Large
\Large\node (A14) [process, below of= A11, yshift=-0cm, xshift= -2.75 cm]{\begin{tabular}{c} \shortstack{Shadow, \eqref{Eq:Shadow} }\end{tabular}};\Large
\Large\node (A17) [process, below of= A11, yshift=-0cm, xshift= 2 cm]{\begin{tabular}{c} \shortstack{\ok{Incident Field}}\end{tabular}};\Large
\Large\node (A13) [process, below of= A11, yshift=-0cm, xshift= -8 cm]{\begin{tabular}{c} \shortstack{Edge-diffraction, \eqref{Eq:EdgeF}}\end{tabular}};\Large
\Large\node (A16) [Solve, below of= A9, yshift=-9cm, xshift= 0 cm]{\begin{tabular}{c} \textsc{\shortstack{Total Fields at a \\detector location $(r,\phi_n)$}}\end{tabular}};\Large
\coordinate (c1) at ([xshift=-4cm] A16.north);
\coordinate (c2) at ([xshift=1.25cm] A16.north);
\coordinate (c3) at ([xshift=2cm] A11.north);
\coordinate (c4) at ([xshift=3.5cm, yshift=-1.5cm] A16.east);
\coordinate (c5) at ([xshift=-6cm, yshift=-0cm] A10.west);
\coordinate (c6) at ([xshift=0.25cm] A15.north);
\coordinate (c7) at ([xshift=-0.25cm] A15.north);

\draw [arrow] (start) -|  (A1);
\draw [arrow] (start) -|  (A3);
\draw [arrow] (A1) --  (A2);
\draw [arrow] (A2) --  (A3);
\draw [arrow] (A3) --  (A4);
\draw [arrow] (A4) --  (A5);
\draw [arrow] (A5) --  (A6);
\draw [arrow] (A6) -- (A7);
\draw [arrow] (A7.east) -|  (A9.north);
\draw [arrow] (A8) |-  (A9);
\draw [arrow] (A9) --  (A10);
\draw [arrow] (A10) -|  (c3);
\draw [arrow] (A11) --  (A12);
\draw [arrow] (A12) -|  (c6);
\draw [arrow] (A15) |-  (A16);
\draw [arrow2] (A9) -|  (c7);
\draw [arrow] (A17) |-  (A16);
\draw [arrow] (A14) --  (c2);
\draw [arrow] (A13) --  (c1);
\draw [arrow] (A12) --  (A13);
%
%\coordinate (c7) at ([xshift= -1cm, yshift=5cm] A10.east);
%\coordinate (c8) at ([xshift=0cm, yshift=0cm] A7.west);     
%\begin{scope}[on background layer]
%\node[draw,dashed,gray,rounded corners,fill=yellow!10, fit=(c8) (A7) (label1)]{};
%\end{scope}

%\begin{scope}[on background layer]
%\node[draw,dashed,gray,rounded corners,fill=yellow!10, fit=(c5) (c4) (label1)]{};
%\end{scope}
%%
\end{tikzpicture}}
\caption{Flowchart illustrating the \red{RO-GSTC} ray tracer for simulating a metasurface specified in terms of its surface susceptibility tensors.}\label{Fig:Flow}
%\end{center}
\end{figure}

\subsection{Numerical Setup}

To test the accuracy of the model, a Forward Ray Tracing (FRT) simulation was developed in MATLAB and compared to an equivalent 2D \sg{Boundary Element Method (BEM) simulation based on Integral Equations Solver}~\cite{smy2020surface, smy2020surface_Camouflage}.
Because of the added complexity required to modify the BEM model for 3D simulations, the ray tracing simulations are also performed in 2D. The main effect on the simulations is the lack of corner-diffracted fields (which are not present in a 2D simulation) as well as the fact that the Keller cone of edge-diffracted scattered rays encompasses the entire plane of the simulation.
% \red{(The previous two sentences may not be necessary now that the entire development is in 2D.)}

\ok{The entire process of synthesizing an electrically large metasurface with certain desired spatial properties, analyzing its surface susceptibility profile, determine $\psi(x)$ from these susceptibilities,  and evaluating its performance in terms of ray-optical fields is illustrated in the flowchart of Fig.~\ref{Fig:Flow}. }

Fig.~\ref{Fig:FRTSetup} shows the simulation setup. In order to analyze the metasurface there are three required components: the surface itself, one or more sources, and multiple field detectors. The metasurface $S$ of length $\ell_x$ is placed in the
%$x-z$
\ok{$x$--$y$}
plane centered on the origin.
%\red{\textbf{[The coordinate system of the metasurface is not consistent. Sometimes in x-y and other times in x-z. Check and possibly fix]}}. \scott{[Only time that it should be/is represented in x-y is in the very first diagram. This could be modified.]} 
Its surface susceptibilities \ok{$\bar{\bm\chi}_{ab}(x)$} are considered known and expressed in the Fourier expansion form of %\eqref{Eq:iFourierSeries}.
\ok{(\ref{Eq:ChiFourierSeries}).}
%In front of the surface, a cylindrical source $\text{Tx}$ is chosen and placed at $(0, -\ell_x)$,
%whose rays interact with the metasurface.
\ok{Cylindrical waves are radiated by a source Tx at $(0, z)$ with $z < 0$, and interact with the surface at $(x,0)$ with $|x| \leq \ell_x/2$.}
To capture the scattering effects as a function of angle, several detectors are placed evenly around the metasurface at locations $\{R, \phi\}$, where each detector has an angular span of $\Delta\phi$.

To analyze the scattered fields from the surface, the
%cylindrical source
\ok{radiated field}
is angularly discretized to generate a set number of
%source
rays (e.g. 1000 rays/$^\circ$) traveling outwards.
%If any of the source \tjs{rays}
\ok{In general, following (\ref{Eq:ScatteredFieldEquations}), when any one of these rays intersects} with $S$, \ok{multiple scattered rays identified by different mode indices $m$ are generated} at the point of intersection.
%relative strength 
%\red{field components}
%and direction
%of the scattered specular rays are computed at the point of intersection
%, obtained
%from the GSTC \red{surface field equations} of Sec.~III-\red{C}.
%In general, at every point of interaction on the surface, multiple \tjs{sets} of scattered rays are generated following \eqref{Eq:ScatteredFieldEquations} identified using the mode index $m$.
This is illustrated using \red{four} incident rays $\{r_1,r_2,r_3,r_4\}$ on the surface in Fig.~\ref{Fig:FRTSetup}, \red{where $r_1$ produced two transmitted rays corresponding to mode index $\{m_0,m_2\}$, while $r_2$ and $r_4$ produced only \tjs{the} \red{first mode  $m_1$} and $r_3$ only the fundamental $m_0$.}
% \red{(The fundamental modes ($m=0$) should follow Snell's law; this is not the case in Fig.~2. I do think it would be instructive to show a fundamental mode.)}
\ok{The field components of these specularly scattered} rays are computed from the GSTC \ok{surface field equations} of Sec.~\ref{sec:IIIC}.

Each of these scattered rays are then further propagated in free space until they pass through a specific detector at $\{R, \phi_n\}$ of $\Delta\phi$ span, which records the
%field strength
\ok{complex electric field vector}
and mode number $m$ of the ray.
The field magnitude and phase of each scattered ray at the detector location are computed using
%the UTD method
\ok{(\ref{Eq:Shadow}) and (\ref{Eq:SpecularF})}
to account for ray propagation \ok{over} their respective distances $d$.
% \red{(It is more common to denote these distances by $s$.)}
For instance, detector $D~\#1$ \ok{in Fig.~\ref{Fig:FRTSetup}} receives the scattering contribution from the rays $\{r_1; m_2\}$ and $\{r_3; m_1\}$. \red{For a general $n^\text{th}$ detector at $\phi = \phi_n$ in the transmission region, the scattered field  $\E_t(R,\phi_n)$ is then the vectorial superposition of all the rays incident on that detector. }
% $I$ ray contributions so that 
% %\scott{[Changed the ray indexing so that it is unique compared to the detector number. May want something better than $i$ for the index.]}
% %
% \begin{align*}
% \E_t(R,\phi_n) = \sum_i^I \red{\sum_m} \E_t^{(m)}(d_i),
% \end{align*}
% %
% where $\E_t^{(m)}(d_i)$ is the complex E-field of the $m^\text{th}$ mode of the $i^\text{th}$ ray obtained by propagating the ray of magnitude $E_t^{(m)}$ by a distance $d_i$.
% \red{(This equation adds little or no value, and suggests incorrectly that one must simply add up all modes. In reality, of course, one only adds those modes of those incident rays that produce a scattered ray that happens to arrive within the span of detector $\#n$.)}
This represents the
%complete
\ok{total specularly scattered field, including the shadow field (\ref{Eq:Shadow}),} %\emph{specular field contributions}
at a specific detector location.

In addition to the specular contributions, rays reaching
%each
\ok{either one} of the two edges of the metasurface at $(\pm \ell_x/2,0)$ \tjs{contribute} to the \emph{edge-diffracted field}. Because the \ok{corresponding} Keller cones span the entirety of the 2D simulation region, the resulting edge-diffracted fields are calculated at the center of each detector. This is illustrated in Fig.~\ref{Fig:FRTSetup} for the \ok{bottom} edge of the surface, where a source ray \red{($r_0$)}  reaches the edge, diffracts, and propagates to detector $D~\#2$. This diffraction contribution is then added to the %contribution from all of the
\ok{specularly scattered field},
%which as a whole
to provide the total scattered field. %across all present detectors.
To obtain the total field,
%in the reflection region,
the field contributions from the incident rays must also be added.
%to the specular \tjs{(including the shadow)} and edge-diffraction fields.
This procedure can be simply extended to obtain the fields in \ok{any} 2D region: instead of just a set of detectors arrayed in a circular formation, an additional set of detectors is arrayed in a 2D grid over the desired simulation region.

% The strength of the specularly diffracted rays are sampled when they pass through the rectangular region of the detector and the rays that are closest to each detector's center is recorded. It should be mentioned that this must be done for each mode, and the contributions from each mode are combined to produce the final field value.

\subsection{Results}

Next, to illustrate the combined \red{RO-GSTC} ray tracing method, several examples will be presented. In all cases, the simulations are performed at a frequency of $f=60$~GHz with a metasurface size of $\ell_x=1$~m (except where noted), i.e., $200\lambda$ in size. To validate the computed field solutions, the results are compared with the BEM-GSTC numerical routine~\cite{smy2020surface, smy2020surface_Camouflage}, which rigorously solves the scattered fields from a metasurface with specified surface susceptibility profile. The BEM simulations are set up to have a mesh density of 20 divisions per wavelength and use the same surface susceptibility function as the equivalent FRT simulation in all cases. Because a $1$~m surface at the chosen frequency is $200\lambda$ in length, it \tjs{is} computationally challenging to use a finer mesh density for the BEM without running into memory limitations. However, in all cases, necessary convergence \tjs{was} ensured.
\subsubsection{\ok{Uniform} Transmissive Surface}
\begin{figure}[h!]
\centering
\begin{overpic}[grid=false, scale = 0.6]{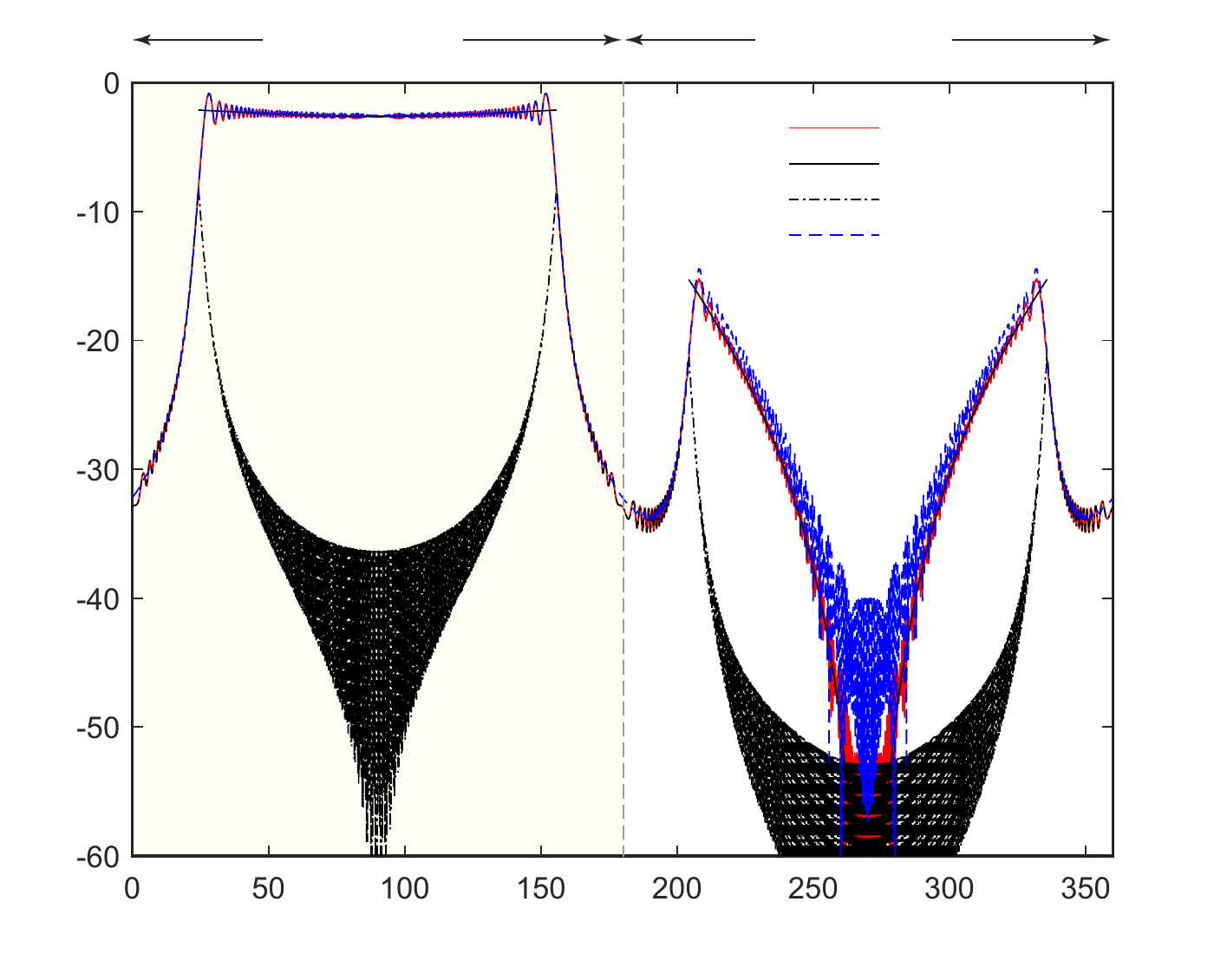}
		\put(74,69.5){\tiny RO-GSTC FRT}
		\put(74,66.55){\tiny Specular}
		\put(74,63.6){\tiny Edge Dif.}
		\put(74,60.65){\tiny BEM-GSTC}
		\put(2,35){\scriptsize \rotatebox{90}{$\|\E_s\|/\|\E_i\|$ (dB)}}
		\put(45,3){\scriptsize $\phi$ (deg)}
		\put(22, 76){\scriptsize \shortstack{ Transmission \\Region,~II}}
		\put(65, 76){\scriptsize \shortstack{ Reflection \\Region,~I}}
	\end{overpic} \caption{\red{Normalized} scattered electric field due to a uniform \red{transmissive} metasurface \red{of size} $\ell_x=1$~m \red{illuminated by a line source at $\ell_z=0.5$~m}.
    The observation distance from the metasurface origin is $R=1~\text{m}$.
	Results computed using \red{RO-GSTC based FRT and BEM-GSTC}.}
	\label{Fig:UniformSurface}
\end{figure}
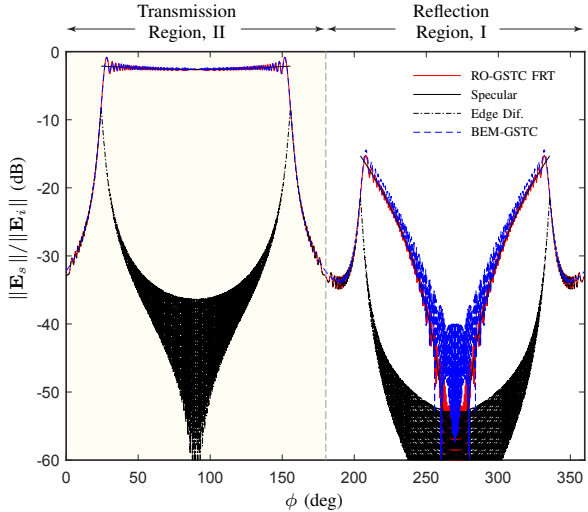

\tjs{The} first example is that of a relatively simple uniform metasurface. For a normally incident plane wave,
%on the metasurface,
the surface susceptibilities are synthesized corresponding to a transmitted plane wave of amplitude $E_t = 0.8j$ and $E_r=0$. The electric and magnetic surface susceptibilities are then computed using \cite{metasurface_synthesis}
\begin{align*}
\chi_\text{ee}(x) &= \frac{2j}{k}\left( \frac{E_t + E_r -1}{E_t + E_r + 1}\right)\\
\chi_\text{mm}(x) &= \frac{2j}{k}\left( \frac{E_t - E_r -1}{E_t - E_r + 1}\right).
\end{align*}
Due to the uniform nature of the metasurface, the parametric variable $\psi$ is constant across the surface \tjs{and} can be used to calculate the susceptibility coefficients using \eqref{Eq:ChiFourierSeries}. Fig.~\ref{Fig:UniformSurface} shows the computed scattered fields for the uniform metasurface for both the FRT as well as the 2D BEM simulations. The FRT results are \tjs{presented as three separate sets of data}: \tjs{1)} the \tjs{contribution} of the critical points of the first kind (Specular); \tjs{2)} \tjs{a contribution from} critical points of the second kind (Edge Dif.); and \tjs{3)} the \tjs{full} combined effect (FRT). These results are normalized with respect to the strength of the incident field at the origin. 

From this plot, we can see a very \tjs{good} agreement between the FRT and BEM simulations, where the fields are almost perfectly overlapped with each other, \tjs{especially} in the transmission region. The interference pattern from the two edge diffraction points \ok{matches} very well for this simple simulation, and the edge-diffracted fields provide a smooth transition across the shadow \ok{boundaries}.
While there is a mild disagreement between the two results in the reflection region ($180^{\ok{\circ}}\leq\phi_s\leq360^{\ok{\circ}}$), the field amplitude is \ok{relatively} small, below $-14$~dB. This disagreement could partially be attributed to the limited mesh density of the BEM-GSTC simulation itself. Finally, it must be noted that even if the surface was synthesized for zero reflection \tjs{given a normally incident planewave}, the actual source in the FRT/\tjs{BEM} simulations is a \ok{cylindrical} source. Therefore \tjs{a} finite amount of reflection \tjs{is} expected.

\subsubsection{Periodically Modulated Metasurface}

For the second comparison, consider a metasurface that has a constant continuous modulation across the entire surface. The parametric variable is set to be \ok{$\psi=0.25x$}, which results in a constant gradient $\ok{\dot \psi=0.25}$ across the surface. For the metasurface synthesis, the transmission fields are chosen to be $E_t^{(m=\pm1)}=0.4j$ and zero for all other values of index $m$, and the reflection is set to be $E_r^{(m)}=0$, both for normal incidence. The susceptibility coefficients are then synthesized using the same method as before and are compared to the equivalent BEM simulation. Because of the large modulation chosen, the length of the metasurface for this simulation was reduced to $\ell_x=0.5$~m; this halving of the surface length allowed the BEM simulation to be \ok{run} with double the mesh density for the same number of mesh elements, thus increasing the accuracy of the BEM simulation for a better comparison. 

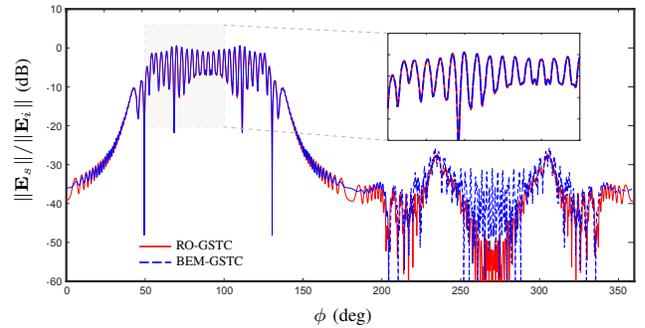
\begin{figure}[h!]
\centering
\begin{overpic}[grid=false, scale = 0.4]{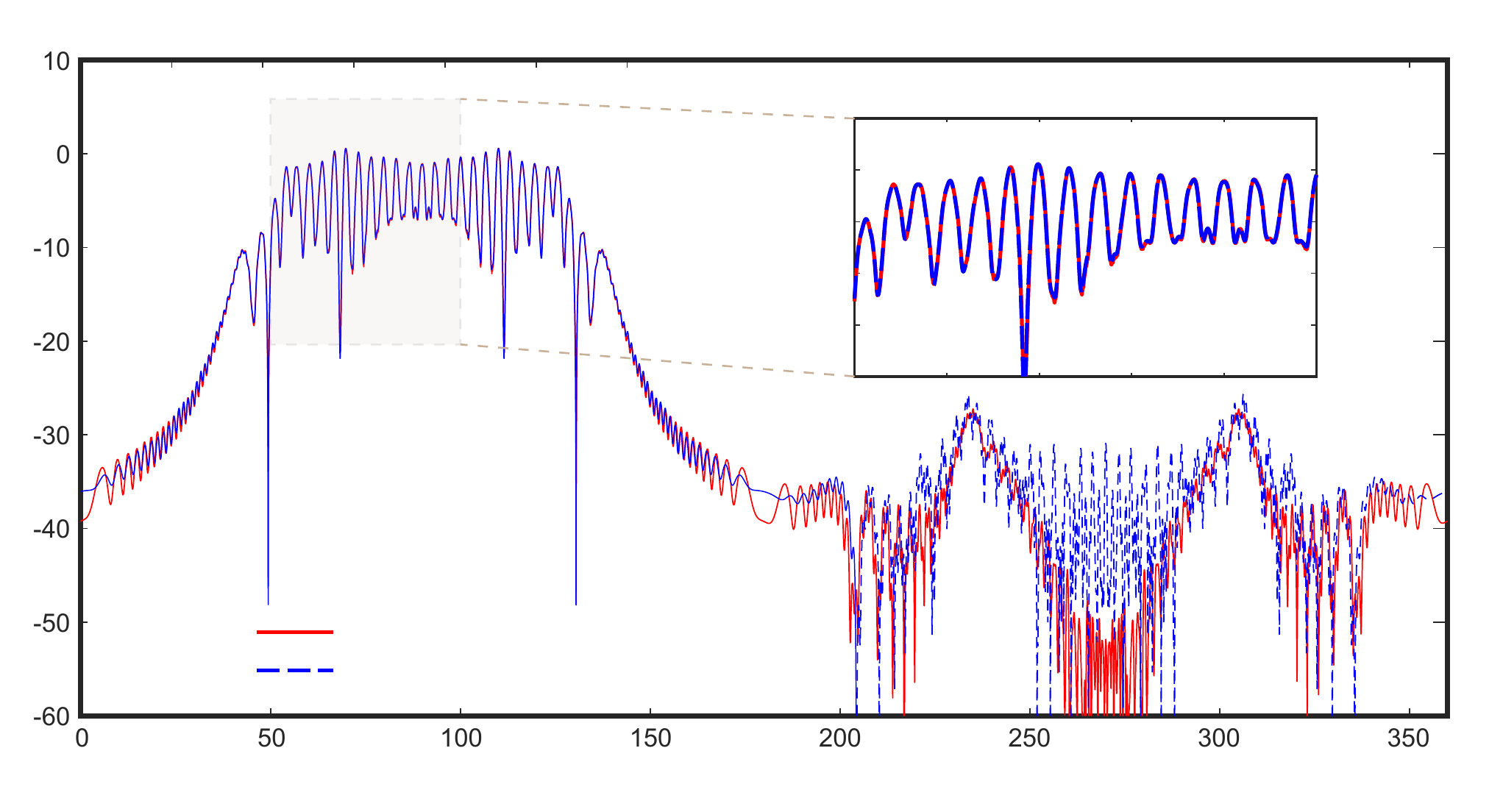}
		\put(23,11.5){\tiny RO-GSTC}
		\put(23,9){\tiny BEM-GSTC}
		\put(-3,19){\scriptsize \rotatebox{90}{$\|\E_s\|/\|\E_i\|$ (dB)}}
		\put(45,0){\scriptsize $\phi$ (deg)}
	\end{overpic}
	\caption{\red{Normalized} scattered \red{electric} field due to a periodically modulated metasurface \red{of size $\ell_x=0.5$~m illuminated by a line source at $\ell_z=0.5$~m.
	The observation distance from the metasurface origin is $R=1$~m.}
	Results computed using RO-GSTC based FRT and BEM-GSTC.
    \red{Inset shows a detail to demonstrate the accuracy of the new method.}}
    %	with a constant period and thus $\ok{\dot \psi} = 0.25$ and $\ell_x=0.5~\text{m}$.}
    \label{Fig:ModulatedSurface}
\end{figure}

\tjs{Figure~}\ref{Fig:ModulatedSurface} compares the simulation results of the scattered fields for both the FRT and BEM for the constant-$\ok{\dot \psi}$ surface. In \tjs{the figure} it can be seen that the two simulations have a strong agreement -- capturing finer details of the diffraction for the transmission region while the BEM becomes very noisy in the reflection region. Because the surface was synthesized for the case of no reflected fields at normal incidence, there is only a very small ($<-25~\text{dB}$) scattered field in the reflection region, making accurate comparisons in this region difficult. 

\subsubsection{Transmissive Beam Diffuser}

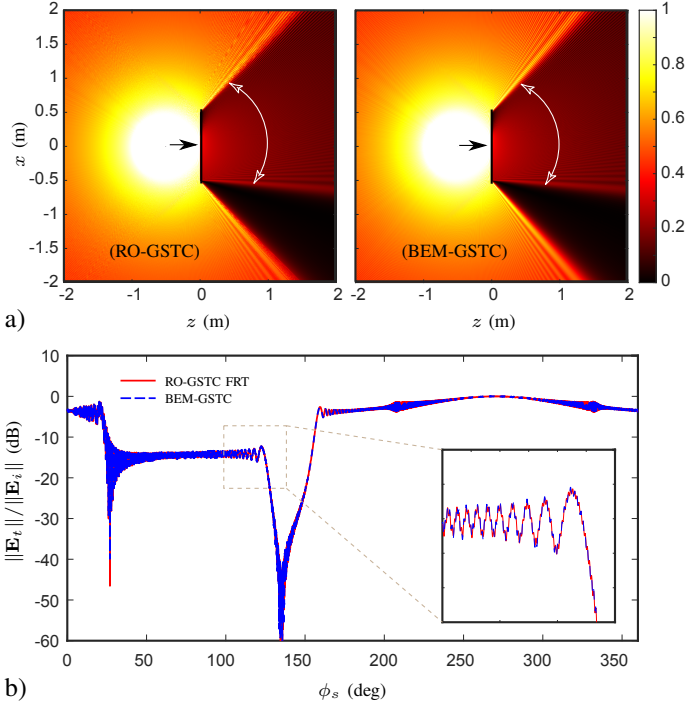
\begin{figure}[h!]
\centering
% \begin{overpic}[grid=false, scale = 0.4]{Figures/SpreaderTotal.pdf}
\begin{overpic}[grid=false, scale = 0.4]{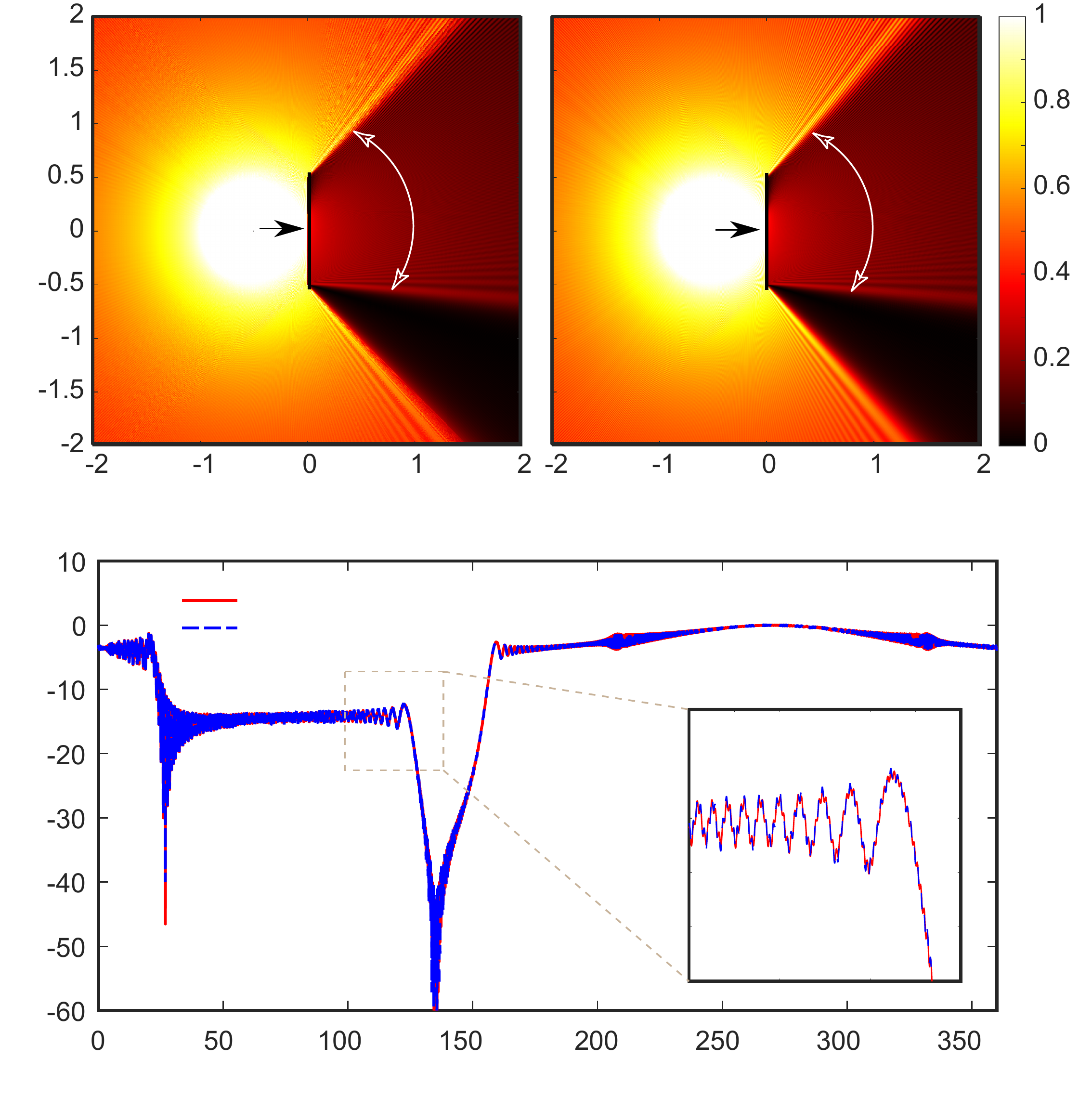}
        \put(0,53){\tjs{a)}}
        \put(0,0){\tjs{b)}}
		\put(23,44.5){\tiny RO-GSTC FRT}
		\put(23, 42){\tiny BEM-GSTC}
		\put(0,19){\scriptsize \rotatebox{90}{$\|\E_t\|/\|\E_i\|$ (dB)}}
		\put(45, 0){\scriptsize $\phi_s$ (deg)}
		\put(26, 53){\scriptsize $z$ (m)}
		\put(67, 53){\scriptsize $z$ (m)}
		\put(57, 63){\scriptsize (BEM-GSTC)}
		\put(15, 63){\scriptsize (RO-GSTC)}
		\put(1, 76){\scriptsize \rotatebox{90}{$x$ (m)}}
	\end{overpic}
	\caption{\red{Normalized} total electric field due to a transmissive diffuser \red{with \mbox{$\ell_x=1$~m}, $\theta_1=35\degree$ and $\theta_2=50\degree$, illuminated by a line source at \mbox{$\ell_z=0.5$~m}:}
	(a) over a 4~m $\times$ 4~m area around the surface, and (b) along an arc of $R=1$~m radius.
	\red{Results computed} using \red{RO-GSTC} based FRT and BEM-GSTC.
	\red{Inset shows a detail to demonstrate the accuracy of the new method.}}
	%with $\theta_1=35\degree$ and $\theta_2=50\degree$. \tjs{Normalized field $|\E_t/\E_i|$ presented:
%	a) over 2D region of $4~\text{m}\times4~\text{m}$ area around the surface and b) along an arc of $R=1~\text{m}$ radius.}
\label{Fig:BeamSpreader}
\end{figure}

The third example that is considered is a transmissive diffuser that diffuses a normally incident plane-wave into a continuous wide range of angles~\cite{yvo,yvo_spreader}. The gradient of the parametric variable can be expressed as
\begin{align}\label{Eq:beamspreader2}
    \ok{\dot \psi}(x) = \frac{\sin\theta_1+\sin\theta_2}{2} + \frac{x}{\ok{\ell_x}}\left(\sin\theta_1-\sin\theta_2\right)
\end{align}
\noindent where $|x|\leq \ell_x$ and $\theta_1$ and $\theta_2$ are the angles of deflection at $x = \pm \ell_x/2,$ respectively. To emulate the behavior of the diffuser, the transmission fields of the first ordered mode is set to \ok{$E_t^{(1)}=0.4j$} and all other transmission coefficients are set to be zero, along with $E_r=0$, to have a perfect matching. Surface susceptibilities are next synthesized and inputted in the \red{RO-GSTC} forward ray tracer. Fig.~\ref{Fig:BeamSpreader} shows the simulation results for this diffuser with $\ell_x=1~\text{m}$, $\theta_1=30\degree$ and $\theta_2=55\degree$, where Fig.~\ref{Fig:BeamSpreader}a shows the 2D total field computed using FRT simulation with a spatial resolution of $\Delta=\lambda/2$, compared with that from BEM simulation. An excellent agreement is seen between the two, even in regions close to the surface. Due to the finite number of rays used in the simulation, the \red{RO-GSTC} 2D field plots are slightly granular compared to finely meshed BEM-GSTC full-wave solver results. The comparison of the 1D total fields along an arc \tjs{(Fig.~\ref{Fig:BeamSpreader}b)} further confirms a close agreement between the two methods.

\subsubsection{Metasurface Collimator}

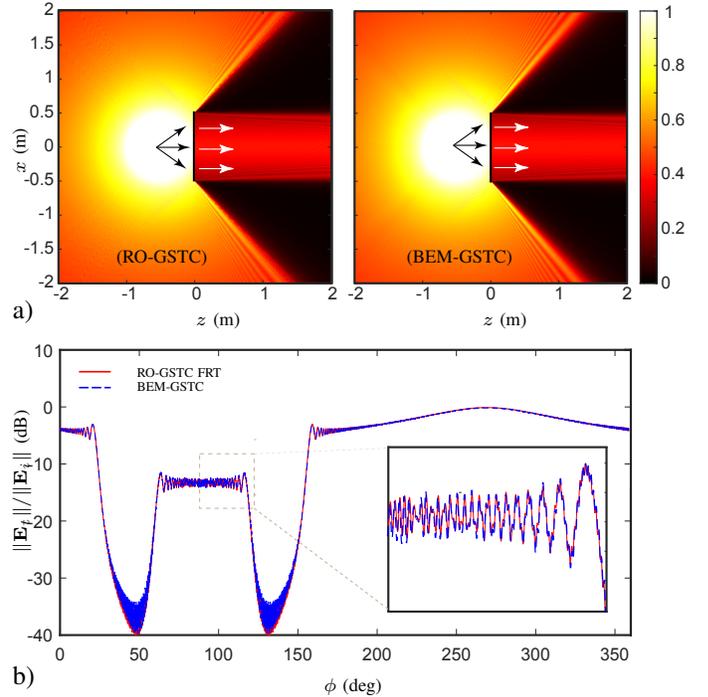
\begin{figure}[h!]
\centering
% \textbf{}\begin{overpic}[grid=false, scale = 0.4]{Figures/FocuserTotal.pdf}
\begin{overpic}[grid=false, scale = 0.4]{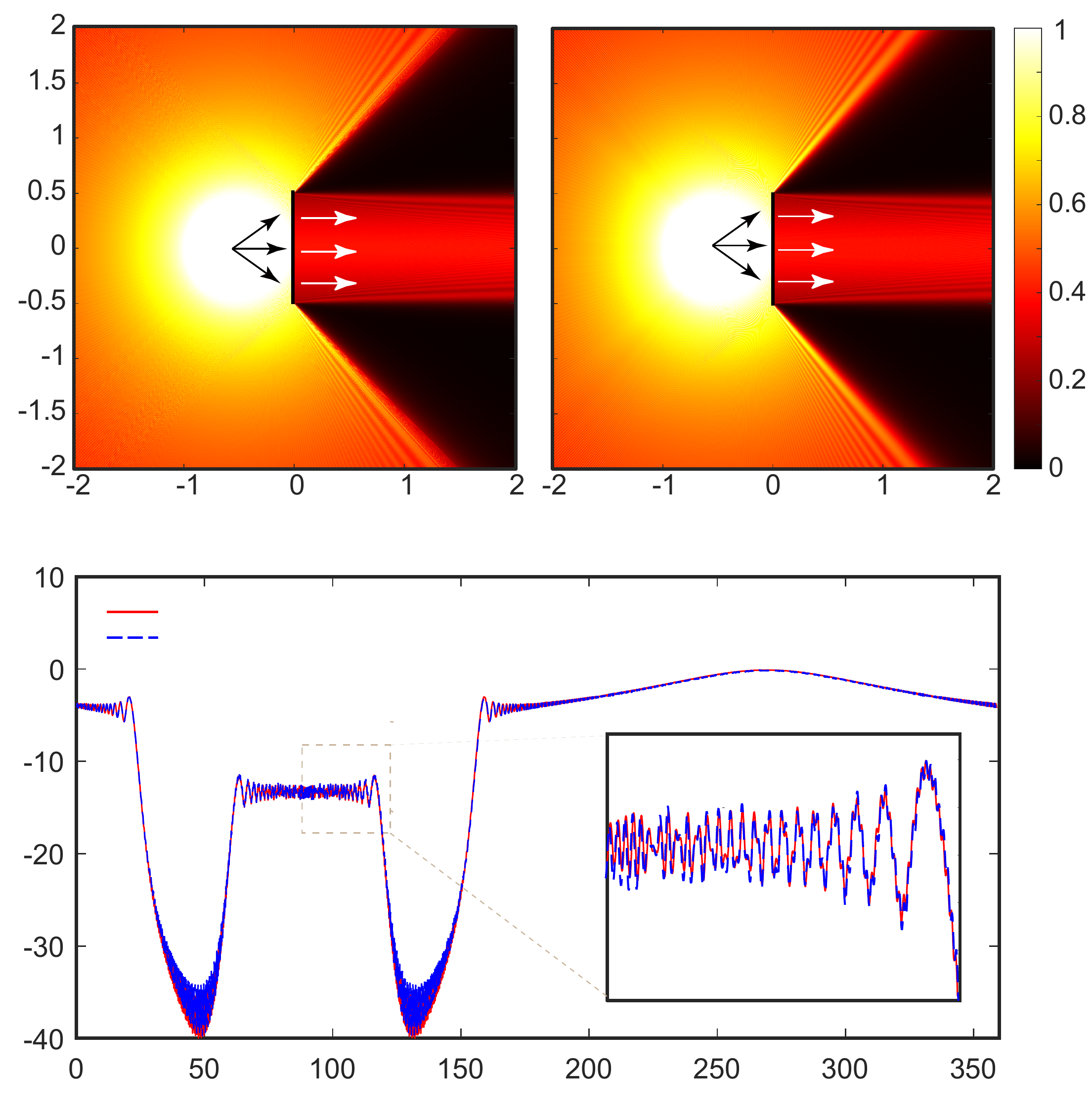}
        \put(0,53){\tjs{a)}}
        \put(0,0){\tjs{b)}}
		\put(18,44.5){\tiny RO-GSTC FRT}
		\put(18,42.5){\tiny BEM-GSTC}
		\put(0,19){\scriptsize \rotatebox{90}{$\|\E_t\|/\|\E_i\|$ (dB)}}
		\put(45, -1){\scriptsize $\phi$ (deg)}
		\put(26.5, 52){\scriptsize $z$ (m)}
		\put(68, 52){\scriptsize $z$ (m)}
		\put(0, 74){\scriptsize \rotatebox{90}{$x$ (m)}}
		\put(57, 61){\scriptsize (BEM-GSTC)}
		\put(15, 61){\scriptsize (RO-GSTC)}
	\end{overpic}
	\caption{\red{Normalized total electric field due to a} metasurface collimator \red{of size $\ell_x=1~\text{m}$, illuminated by a line source at $\ell_z=0.5$~m: (a) over a 4~m $\times$ 4~m area around the surface, and (b) along an arc of $R=1$~m.}
	\red{Results computed using RO-GSTC based FRT and BEM-GSTC.}
%	Shows the total \tjs{normalized field $|\E_t/\E_i|$ for: a)} the area around the metasurface caused by a cylindrical point source at $(0, -0.5)$m computed using Forward Ray Tracing (FRT) \tjs{and BEM-GSTC; and (b) } the extracted total fields along the arc of a circle of radius $R=1$m for both FRT and the BEM-GSTC
    \red{Inset shows} a detail to demonstrate the accuracy of the new method.}
\label{Fig:FocuserResults}
\end{figure}

The final example is \tjs{the} metasurface collimator described in Sec.~IV. This metasurface acts as a lens which, when excited with a cylindrical source at its focal point, can collimate the beam on the transmission side, generating an outgoing plane-wave normal to the surface. The extracted surface susceptibilities have already been shown in Fig.~\ref{Fig:SSFT_Exp}, and its resulting Fourier expansion form with the extracted $\psi$ and $\ok{\dot \psi}$ is next used in the following simulation results. Fig.~\ref{Fig:FocuserResults} shows the total field simulation results for the metasurface collimator comparing the FRT and BEM simulations. As expected, the point source is collimated in transmission with negligible reflection. An excellent agreement is seen between the two simulations even at low amplitude levels, where the interference patterns are near identical between the two methods. Again due to the finite number of rays used in the simulation, the \red{RO-GSTC} 2D field plots are granular compared to finely meshed BEM-GSTC full-wave solver results.

\tjs{The  variety of examples presented above} demonstrate the usefulness of the proposed \red{RO-GSTC} forward ray tracer \tjs{by which} the \tjs{generated} scattered fields can be efficiently \tjs{modeled for} metasurfaces described in terms of their \tjs{constituent surface parameters}.

\section{Conclusion}
This work has continued from the
\ok{ray-optical method in \cite{yvo} for computing the scattered fields from finite, locally periodic metasurfaces
characterized by their reflection and transmission coefficients $\{{\bm\Gamma,{\bf T}}\}$.}
%UTD ray tracing
%method based on transmission and reflection coefficients $\{\T, \mathbf{\Gamma}\}$ developed for computing the scattered fields from metasurface in \cite{yvo}.
\ok{Recognizing that $\{{\bm\Gamma,{\bf T}}\}$ are angle-dependent, the method proposed herein} captures the metasurface scattering behavior using their field-independent constitutive parameters, i.e., angle-independent surface susceptibilities. \sg{This leads to compact and a physically motivated description of the surface, while still being able to correctly capture its macroscopic angular response through its tangential and possible normal susceptibility components, if they exist in their corresponding physical implementations}. For slowly varying non-uniform metasurfaces, the surface susceptibilities have been expressed in \tjs{a} Fourier expansion form and integrated into the GSTCs to solve for the scattered \tjs{transmitted and reflected} fields.
%Coupled with
\ok{Making use of a forward ray tracing method} based on GO/UTD, the scattered \ok{surface} fields
%have been
are propagated to the desired observation regions, enabling the total scattered field \ok{to be constructed} anywhere. The resulting \red{RO-GSTC} framework has been verified \ok{for a variety of examples} using a BEM-GSTC full-wave solver. As long as the specified susceptibility profile can be expressed in the Fourier expansion form using the parametric variable $\psi$, the proposed method can solve for the scattered fields. It thus represents a powerful and efficient numerical platform to solve electrically large metasurface problems at the cost of sacrificing \ok{some of} \tjs{the accuracy of a full-wave solution}.

\bibliographystyle{IEEEtran}
\bibliography{references.bib}

\end{document}